\documentclass[review]{elsarticle}

\journal{Medical Engineering and Physics}

\usepackage{lineno}
\usepackage{hyperref}

\usepackage{graphicx}
\usepackage{caption}
\usepackage{subcaption}
\usepackage{breqn} 
\usepackage{bm}	
\usepackage{booktabs}	
\usepackage{multirow} 
\usepackage[table,xcdraw]{xcolor} 

\begin{document}

\begin{frontmatter}
	
	\title{Evaluation of Matrix Factorisation Approaches for Muscle Synergy Extraction}
	
	\author[Edi]{Ahmed Ebied \corref{correspondAuthor}}
	\ead{ahmed.ebied@ed.ac.uk}
	\author[Edi]{Eli Kinney-Lang}
	\author[Edi]{Loukianos Spyrou}
	\author[Edi]{Javier Escudero}
	\address[Edi]{School of Engineering, Institute for Digital Communications, The University of Edinburgh, Edinburgh EH9 3FB, United Kingdom}


	\cortext[correspondAuthor]{Corresponding author}
	\tnotetext[t1]{\textbf{Word count:} 4402}

\begin{abstract}
The muscle synergy concept provides a widely-accepted paradigm to break down the complexity of motor control. In order to identify the synergies, different matrix factorisation techniques have been used in a repertoire of fields such as prosthesis control and biomechanical and clinical studies. However, the relevance of these matrix factorisation techniques is still open for discussion since there is no ground truth for the underlying synergies. 
Here, we evaluate factorisation techniques and investigate the factors that affect the quality of estimated synergies. We compared commonly used matrix factorisation methods: Principal component analysis (PCA), Independent component analysis (ICA), Non-negative matrix factorization (NMF) and second-order blind identification (SOBI). Publicly available real data were used to assess the synergies extracted by each factorisation method in the classification of wrist movements. Synthetic datasets were utilised to explore the effect of muscle synergy sparsity, level of noise and number of channels on the extracted synergies.
Results suggest that the sparse synergy model and a higher number of channels would result in better estimated synergies. Without dimensionality reduction, SOBI showed better results than other factorisation methods. This suggests that SOBI would be an alternative when a limited number of electrodes is available but its performance was still poor in that case. Otherwise, NMF had the best performance when the number of channels was higher than the number of synergies. Therefore, NMF would be the best method for muscle synergy extraction.

\end{abstract}
	
	
\end{frontmatter}


\section{Introduction}

\subsection{Muscle synergy}
``How does the central nervous system (CNS) control body movements and posture?'' This question has been discussed for over a century with no conclusive answer. The coordination of muscles and joints that accompanies movement requires multiple degree of freedoms (DoFs). This results a high level of complexity and dimensionality \cite{DAvella2015}. A possible explanation to this problem considers the notion that the CNS constructs a movement as a combination of small groups of muscles (synergies) that act in harmony with each other, thus reducing the dimensionality of the problem. This idea could be traced to the first decades of the twentieth century \cite{Sherrington1910a} and has been formulated and developed through the years \cite{Bizzi1991,Giszter1993,Mussa-Ivaldi1994} to reach the Muscle Synergy hypothesis \cite{Tresch1999,Saltiel2001,DAvella2003}. The muscle synergy concept posits that the CNS achieves any motor control task using a few synergies combined together, rather than controlling individual muscles. Although the muscle synergy hypothesis is criticized for being very hard to be falsified \cite{TRESCH}, a repertoire of studies have provided evidence and support for it. Those pieces of research could be categorized into two main categories: direct stimulation and behavioural studies.

The stimulation approaches were conducted by exciting the CNS at different locations to study the resulting activation pattern. Earlier studies focused on the organization of motor responses evoked by micro-stimulation of the spinal cord of different vertebral species, such as frogs \cite{Bizzi1991,Giszter1993,Mussa-Ivaldi1994,Kargo2008,Hart2004}, rats \cite{Tresch1999a} and cats \cite{Lemay2004}. They revealed that the responses induced by simultaneous stimulation of different loci in the spinal cord are linear combinations of those induced by separate stimulation of the individual locus. Those findings were supported by another direct stimulation studies where a relatively long period of electric stimulation applied to different sites in the motor cortex resulted in complex movements in rats \cite{Haiss2005}, prosimians \cite{Stepniewska2005} and macaques \cite{Overduin2008,Overduin2014a}. The chemical micro-stimulation has been used through N-methyl-D-aspartate iontophoresis injected into the spinal cord of frogs which evoked an electromyographic (EMG) patterns that could be constructed as a linear combination of a smaller group of muscle synergies \cite{Saltiel2001}.

Similarly, the behavioural studies rely on recording the electrical activity of the muscles (electromyogram, EMG) during a specific task (or tasks) or natural behaviour. Then, a number of synergies is extracted from the signals using computational techniques. The identified synergies should be able to describe the recorded signal for the related task or behaviour. Studies have been carried out on cats where four muscle synergies were sufficient to reproduce 95\% of postural hind-limb muscles response data \cite{Ting2005} and five synergies accounted for 80\% of total variability in the data \cite{Torres-Oviedo2006}. Similar research on monkeys during  grasping activity showed that three muscle synergies accounted for 81\% of variability \cite{Overduin2008}. In humans, muscle synergies were identified from a range of motor behaviours \cite{Cheung2005,Weiss2004} with the ability to describe most of the variability in EMG signals. In addition, other studies show that complex motor outputs such as upper limb reaching movements \cite{DAvella2006}, cycling \cite{Wakeling2009,Hug2011a} and  human postural control \cite{Ting2007} are a result of the combination of few muscle synergies.

In the recent years, many studies applied the muscle synergy concept to analyse and study body movements and muscle coordination in diverse applications. For instance, it has been used to establish the neuromuscular system model \cite{Aoi2015}. Moreover, the hypothesis has been used in many clinical applications \cite{Pons2016a} in addition to several biomechanical studies such as walking and cycling \cite{Nazifi2017,Martino2015}. The extracted synergies are utilised in prosthesis control through classification \cite{Rasool2016,Ma2015a} and regression \cite{Jiang2014b}. 

\subsection{Mathematical models for muscle synergy}

In all studies, the muscle synergies are estimated from the recorded electrical activity of the muscle. Signals are either collected using surface EMG or invasively using needle EMG. Then, the EMG recordings needs to be modelled in order to compute the muscle synergies.

Two main muscle synergy models have been proposed: the time invariant or synchronous model \cite{Tresch1999,Saltiel2001} and the time-varying or asynchronous model \cite{DAvella2002,DAvella2003}. The electrical activity for single muscle or channel $\textbf{m}(t)$  is a vector that could be expressed according to the time-invariant model as a combination of synchronous synergies $\textbf{s}$ (scalar values activated at the same time) multiplied by a set of time-varying coefficients or weighting functions $\textbf{w}$ as shown in equation \ref{eq_time_invarient}
\begin{equation}\label{eq_time_invarient}
\textbf{m}(t)=\sum_{i=1}^{i=r}s_i \textbf{w}_i(t)
\end{equation}
where $r$ is the number of synchronous synergies. Since synergies contribute to each muscle activity pattern with the same weighting function $\textbf{w}_i(t)$, the synergy model is synchronous without any time variation.

On the other hand, the time-varying synergies are asynchronous as they are compromised by a collection of scaled and shifted waveforms, each one of them specific for a muscle or channel. Thus, the muscle activity  $\textbf{m}(t)$ can be described according to the asynchronous model with a group of time-varying synergy vectors scaled and shifted in time by $c$ and $\tau$, respectively, as shown in equation \ref{eq_time_varying}.
\begin{equation}\label{eq_time_varying}
\textbf{m}(t)=\sum_{i=1}^{i=r}c_i \textbf{s}_i(t-\tau_i)
\end{equation}
In this case, the model is capable of capturing fixed relationships among the muscle activation waveforms across muscles and time. By means of comparison, time-invariant synergies can acquire the spatial structure in the patterns but any fixed temporal relationship can be recovered only indirectly from the weighting functions associated with its synchronous synergy.

Although the time-varying model provides a more parsimonious representation of the muscle activity compared to the time-invariant model, some studies have shown evidence that the muscle synergies are synchronised in time \cite{Hart2013,Kargo2008}. Therefore, most recent muscle synergies studies apply the time-invariant model for synergy extraction. This is done by using matrix factorization techniques on multichannel EMG activity to estimate the muscle synergies and their weighting functions.

\subsection{Comparison of Matrix factorization techniques}

According to the time-invariant model, the estimation of muscle synergies (spatial profile) and their weighting functions (temporal profile) from a multi-channel EMG signal is a blind source separation (BSS) problem. This problem is approached by matrix factorisation techniques to estimate the set of basis vectors (synergies). Various matrix factorisation algorithms have been applied based on different constraints. The most commonly used factorisation techniques to extract synergies for myoelectric control and clinical purposes are principal component analysis (PCA) \cite{Jackson1991} which was applied in \cite{Ranganathan2012}, independent component analysis (ICA) \cite{Hyvarinen2000} that was used in \cite{Rasool2016} and \cite{Kargo2003}, in addition to non-negative matrix factorization (NMF) \cite{Lee1999} which have been used in \cite{Choi2011,Jiang2014b} and \cite{Berger2014}.

In this paper, these three techniques are compared among themselves and to second-order blind identification (SOBI) \cite{Belouchrani1997}, a technique which has not been used for muscle synergy estimation previously. A first evaluation of the matrix factorisation algorithms for muscle synergy extraction was reported in 2006 \cite{Tresch2006} where the algorithms were tested with simulated data under different levels and kinds of noise and they were applied on real data to show the similarities between their estimated synergies. A more recent study \cite{Lambert-Shirzad2017} used joint motion data to evaluate kinematics and muscle synergies estimated by PCA, ICA and NMF using the quality of reconstructing the data by synergies as a metric for evaluation.
Here, we are concerned with nature and number of muscle synergies and the factors that affect their quality which have not been discussed by those studies. The sparsity of synergies is investigated where synthetic sparse and non-sparse synergies are compared to study their effect on the matrix factorisations. Moreover, the ratio between number of channels and synergies (dimension reduction ratio) is studied. Those comparisons are carried out under different noise levels to show the robustness of factorisation methods to noise. In addition, synergies extracted from a real dataset by the four matrix factorisation techniques were used to classify between wrist movements. The classification accuracy was used as a metric in the factorisation methods comparison. We aim to compare current matrix factorisation techniques in addition to SOBI and investigate the factors that affect the quality of their extracted synergies such as sparsity and channel/synergy ratio.

\section{Methods}

\subsection{Real dataset}

\begin{figure}[]
	\centering
	\includegraphics[width=1\linewidth]{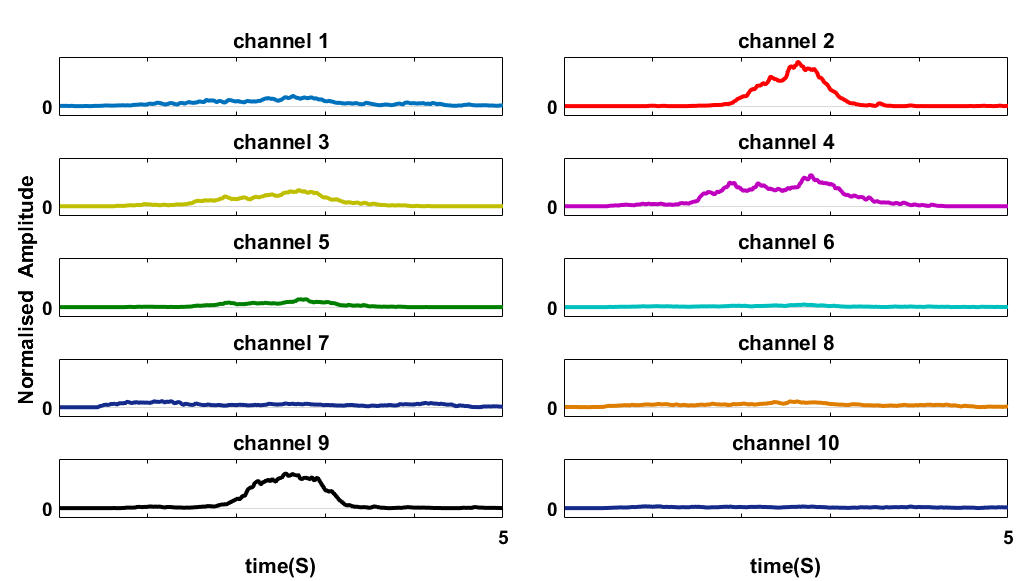}
	\caption{Example of 10-channel EMG envelopes recorded during wrist extension movement for 5 s of Subject 4/repetition 1 (the amplitude is normalised only in figure to highlight the differences between channels).}
	\label{fig:subplotemg}
\end{figure}

We used the Ninapro first dataset \cite{Atzori2014,Atzori2015a} which consists of recordings for 53 wrist, hand and finger movements. Each movement/task has 10 repetitions from 27 healthy subjects. The dataset contains 10-channel signals rectified by root mean square and sampled at 100 Hz as shown in Figure \ref{fig:subplotemg}. The real dataset is used in the comparison between matrix factorisation techniques. Moreover, it is used as a part of the synthetic data creation as discussed in \ref{subSec_symthetic_data}.

For the real data comparison, the three main degree of freedoms (DoF) investigated for the wrist motion are wrist flexion and extension (DoF1),  wrist radial and ulnar deviation (DoF2), and  wrist supination and pronation (DoF3). Wrist movement through these three degrees of freedom are essential  for prosthetic control \cite{Farina2014c}. Thus, they may highlight the application of muscle synergies in myoelectric control.

\subsection{Synthetic data}\label{subSec_symthetic_data}

The performance of each matrix factorisation algorithm was tested using synthetic datasets as ground truth. Since the studies  \cite{Hart2013,Kargo2008} showed an evidence that the muscle synergies are synchronised in time, the data was generated according to the time-invariant model \cite{Tresch1999} in which EMG activity for  $j^{th}$-channel is the summation of its coefficients in each synergy ($s_{ij}$), weighted by the respective weighting function ($\textbf{w}_i(t)$), as the following:
\begin{equation}\label{eq_time_inv}
\textbf{m}_{j}(t)=\sum_{i=1}^{i=r}s_{ij}\textbf{w}_i(t)+g(\epsilon)
\end{equation}
where $\textbf{m}_{j}(t)$ is the simulated EMG data over channel $j$, while  $\epsilon$ is a Gaussian noise vector and $g(x)$ is the Heaviside function used to enforce non-negativity. For $m$-channel data, this model could be expanded into its matrix form. In this case, the synthetic EMG data $\mathbf{M}$ is a matrix with dimensions (\textit{m} channels$\times$\textit{n} samples) as
\begin{equation}\label{eq_matrix_form}
\mathbf{M}_{(m\times n)}  = \mathbf{S}_{(m\times r)} \times \textbf{W}_{(r\times n)} + g(\textbf{E})
\end{equation}
where $r$ is the number of synergies ($r$\textless$m$) and \textbf{E} is the matrix form of the Gaussian noise vector $\epsilon$ for all channels. $\mathbf{S}$  $(m\times r)$ and $\textbf{W}$  $(r\times n)$ are the synergy matrix and weighting function matrix form, respectively.

\begin{figure*}[t]
	\centering
	\begin{subfigure}[b]{0.49\textwidth}
		\centering
		\includegraphics[width=\textwidth]{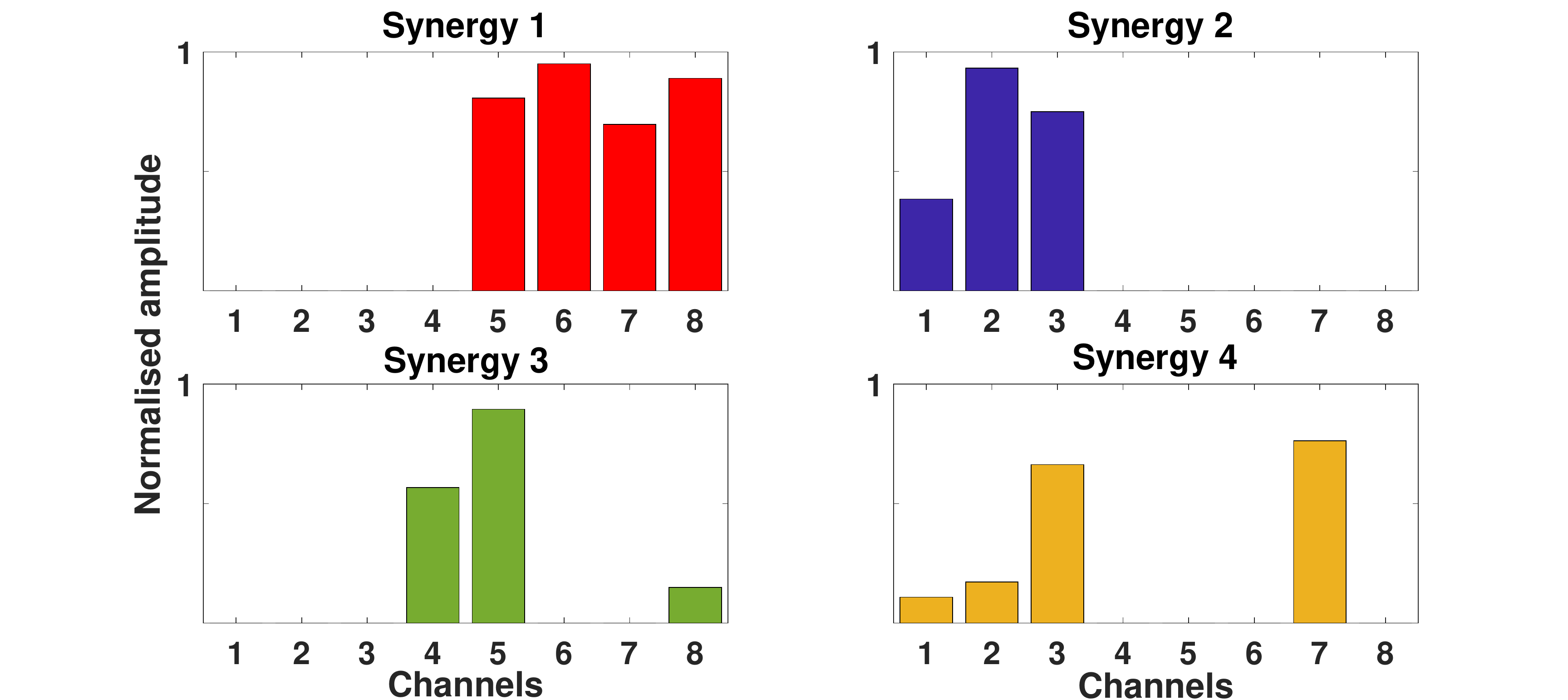}
		\caption[]%
		{{\small Synthetic sparse synergies }}    
		\label{fig:ExampleSparseSyn}
	\end{subfigure}
	\hfill
	\begin{subfigure}[b]{0.49\textwidth}  
		\centering 
		\includegraphics[width=\textwidth]{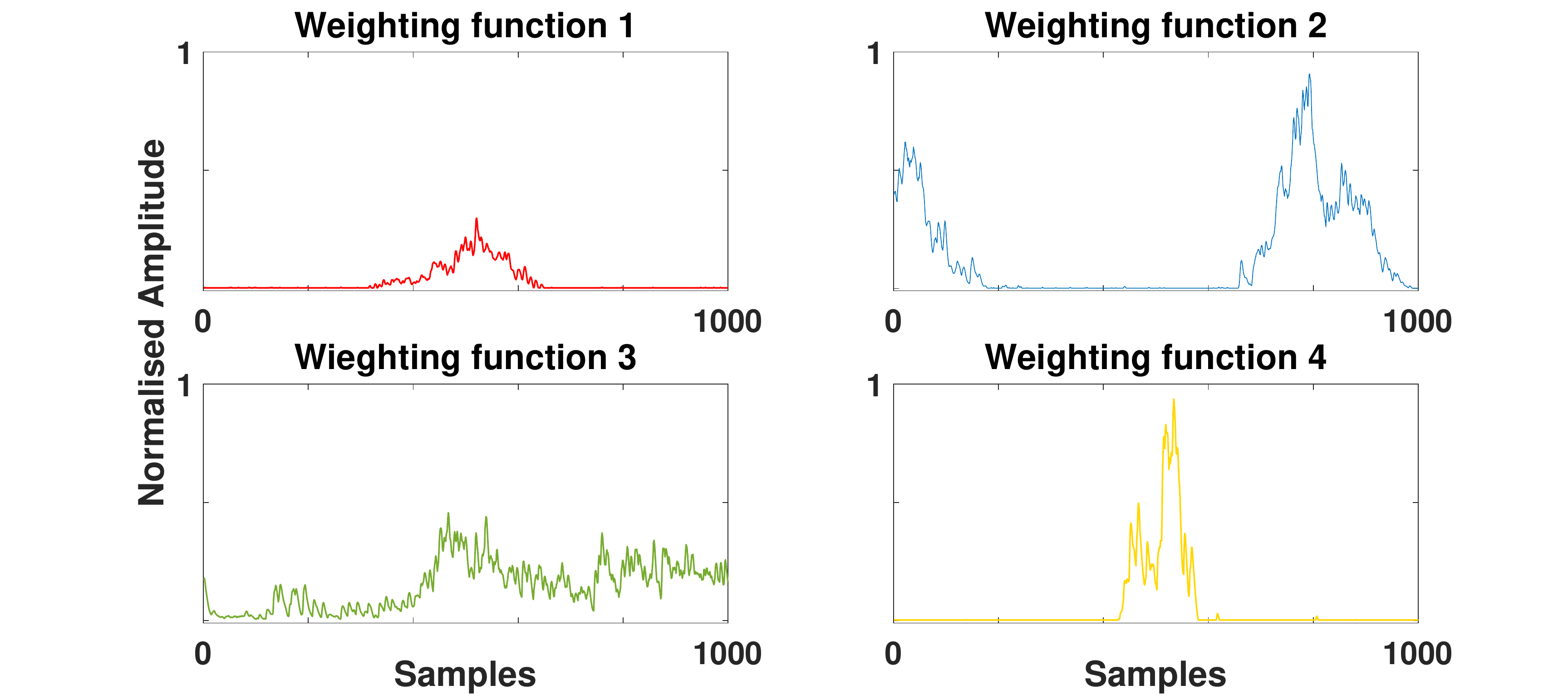}
		\caption[]%
		{{\small Weighting functions}}    
		\label{fig:ExampleSparseWeight}
	\end{subfigure}
	\vskip\baselineskip
	\begin{subfigure}[b]{0.8\textwidth}   
		\centering 
		\includegraphics[width=\textwidth]{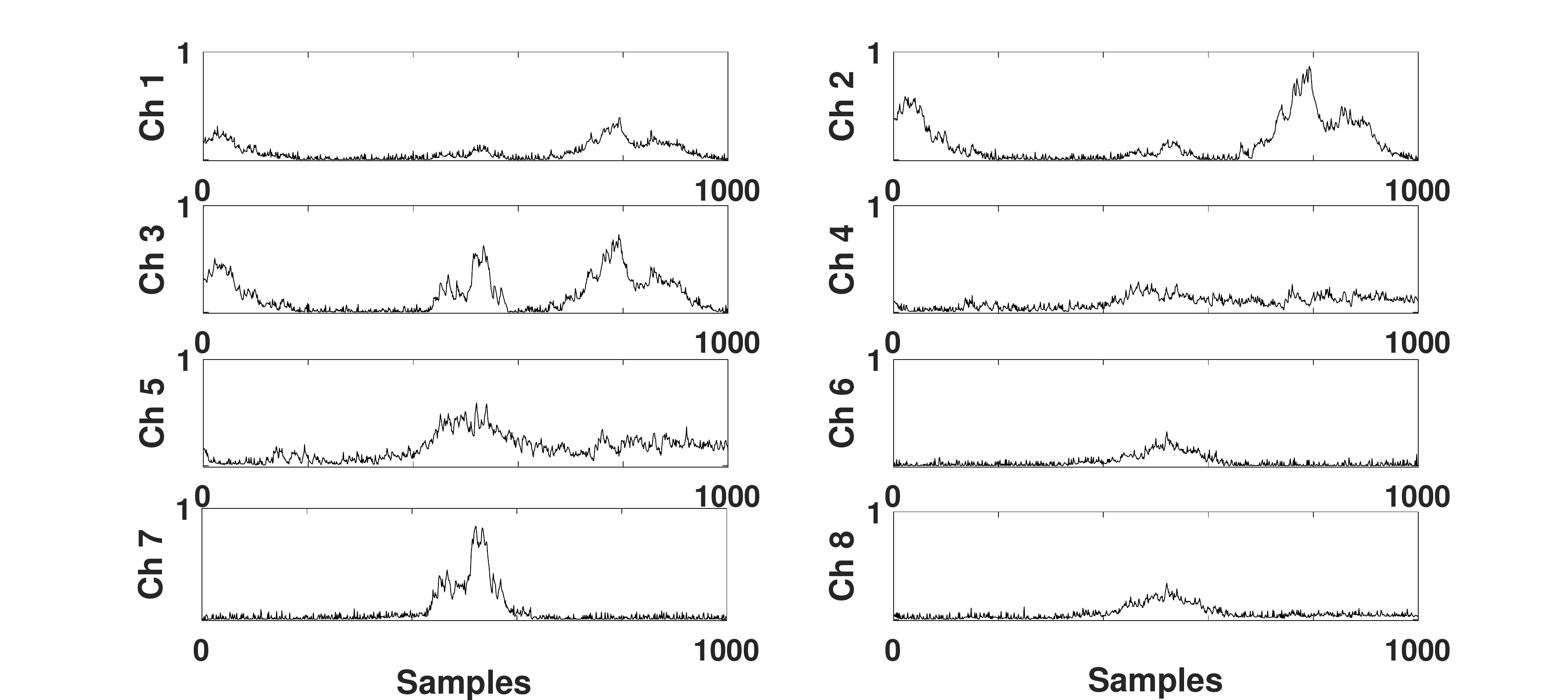}
		\caption[]%
		{{\small The resulting synthetic EMG dataset (after adding the noise). }}    
		\label{fig:ExampleSparseSig}
	\end{subfigure}

	\caption{An example of 8-channel synthetic EMG signal (Panel \ref{fig:ExampleSparseSig}) creation using four sparse synergies (Panel\ref{fig:ExampleSparseSyn}) and their respecting weighting functions (Panel \ref{fig:ExampleSparseWeight}) which is a randomly selected real EMG segments with  15 dB SNR.}
	
	\label{fig:ExampleSparse}
\end{figure*}

In order to generate a synthetic EMG signal that mimics the real EMG data and carries the synergistic information, the three elements in equation \ref{eq_matrix_form} should be designed so that they reflect real activities under diverse assumptions. The synergy matrix $\mathbf{S}_{(m\times r)}$ was assigned a non-negative random values between [0,1] to retain the additive nature of synergies, while each weighting (activation) function $\textbf{W}_{(r\times n)}$ is a real EMG envelope randomly assigned from the Ninapro dataset from different subjects and movements. This approach based on real data was chosen to ensure that the generated signal retains the statistical properties of the EMG signal rather than assigning randomly generated signals for the weighting function as done in the past \cite{Tresch2006}. Finally, the non-negative part of the Gaussian noise is applied to the mixture by the Heaviside function $g(\textbf{E})$. An example of the generated synthetic EMG signal is shown in Figure \ref{fig:ExampleSparse}.

The synthetic signals were generated with different settings to compare the factorisation methods under various conditions. In all settings, the number of synergies ($r$) was fixed to four synergies. This choice was based on the fact that the number of synergies used in previous studies varied from one or two synergies \cite{Jiang2014b} to six synergies \cite{Torres-Oviedo2007}, for example.

Three criteria were investigated: the sparsity of synergy matrix, the number of channels, and the added noise level. The sparsity of the synergy matrix  $\mathbf{S}_{(m\times r)}$  is investigated since all muscles (channels) may be not activated during a specific movement at the same time. The sparse synergies were created by constraining each channel by 40\% sparsity level (i.e., a maximum of for channels being active in each synergy) to ensure that each channel has at least one non-zero value in the four synergies. This approach would typically avoid having channels that are inactive in all 4 synergies as shown in  Figure \ref{fig:ExampleSparseSyn} as an example of sparse synthetic synergies. In comparison, the non-sparse synergies are non-negative random values between [0,1]. Secondly, the effect of dimension reduction between the generated signal and synergies (basis vectors) is examined. The number of synergies is fixed to 4 in all settings while the number of channels are 4 (no dimension reduction), 8 or 12 channels. Finally, the effect of additive signal to noise ratio (SNR) is compared at three levels: 10, 15 and 20 dB. In total, 10 synthetic datasets are generated, each containing 1000 separate trials for each setting.


\subsection{Matrix factorisation algorithms}

The muscle synergy time-invariant model is approached as a blind source separation problem, where a multichannel EMG signal matrix $\textbf{M}(t)$ is modelled as a linear mixture of synergies and ``source signals''. Therefore, according to equation \ref{eq_time_invarient}, $\textbf{M}(t)$ will follow the linear matrix factorisation model as follows
\begin{equation}\label{eq_factorisation_model}
\textbf{M}(t)=\textbf{S}\textbf{W}(t)
\end{equation}
In this context, $\textbf{S}$ is the mixing (synergy) matrix  while $\textbf{W}(t)$ contains the source vectors (weighting functions) with dimensions number of synergies $\times$ time. The noise is disregarded in equation \ref{eq_factorisation_model}. In order to estimate unique solutions, additional constraints are needed.
 
PCA constrains the components of the model in equation \ref{eq_factorisation_model} to be orthogonal, where the first component holds the largest variance and the variance progressively decreases for each component \cite{Abdi2010}. Here, PCA has been performed using the``pca'' Matlab function (version 2016a).

For ICA, the fixed-point algorithm introduced in \cite{Hyviirinen} has been used. Unlike PCA, ICA attempts to extract independent components by whitening the data to remove any correlation. Then, it rotates the pre-whitened data to extract the non-Gaussian components.

NMF imposes a non-negative constraint on the extracted factors. The algorithm relies on a cost function to quantify the quality of approximation between the data matrix $\textbf{M}$ and its factorised non-negative matrices $\textbf{S}$ and $\textbf{W}$ where $\textbf{M}\approx \textbf{SW}$. Values of $\textbf{S}$ and $\textbf{W}$ are updated and optimised to find the local minima numerically. The Matlab function "nnmf" (version 2016a) was used to perform the NMF based on \cite{Berry2007}. 

SOBI \cite{Belouchrani1997} has not been applied to extract muscle synergies before. However, it is included in this comparison because SOBI utilises the joined diagonalisation of time delayed covariance matrices to estimate the unknown components. Therefore, it could reveal more information about the temporal profile of the EMG activity. Thus, SOBI leads to components that are uncorrelated at those time delays and, therefore, it is sometimes considered an alternative to ICA, which is based on higher order statistics. Here, SOBI was performed using the default 4 diagonalised covariance matrices with the function "sobi" in the ICALAB package \cite{Tanaka}.
 
\begin{figure*}[t]
	\centering
	\begin{subfigure}[b]{0.475\textwidth}
		\centering
		\includegraphics[width=\textwidth]{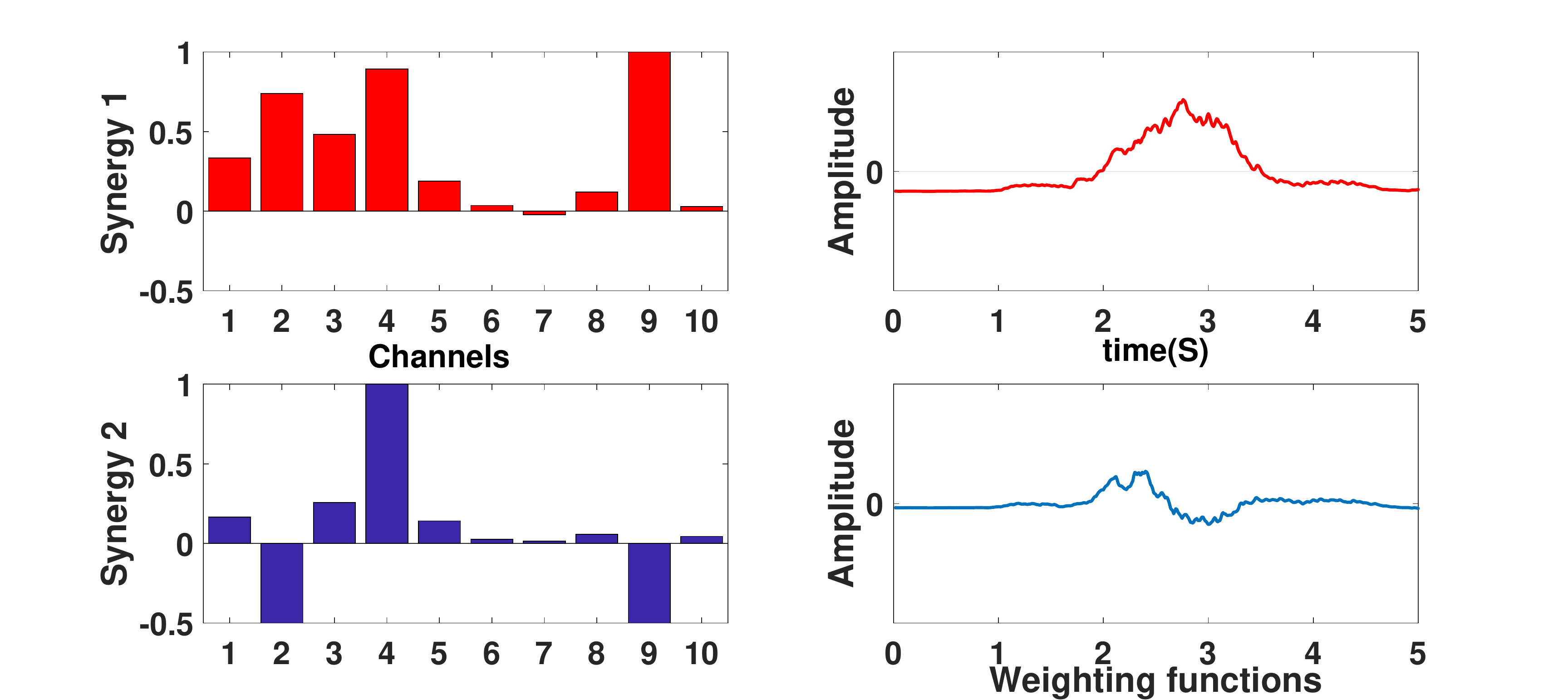}
		\caption[]%
		{{\small PCA synergies}}    
		\label{fig:synergyPCA}
	\end{subfigure}
	\hfill
	\begin{subfigure}[b]{0.475\textwidth}  
		\centering 
		\includegraphics[width=\textwidth]{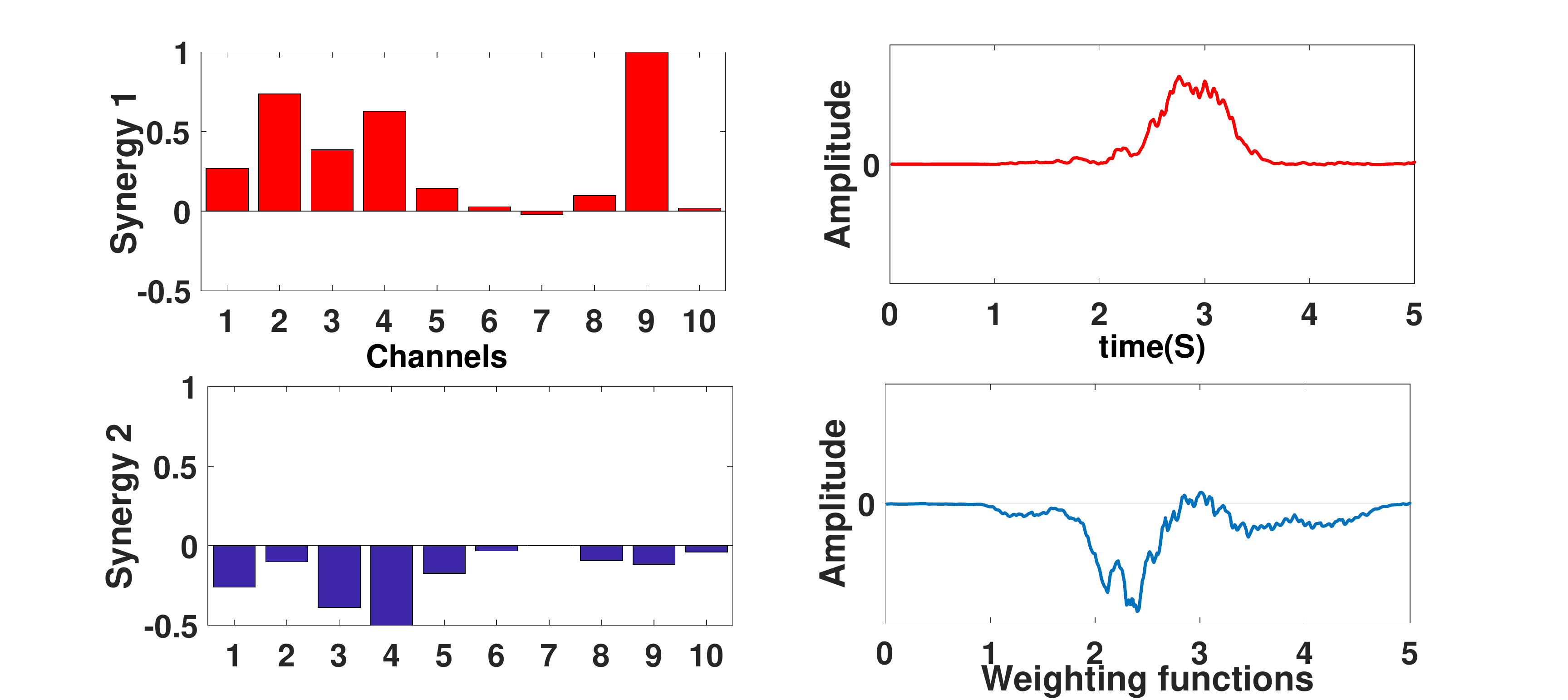}
		\caption[]%
		{{\small ICA Synergies}}    
		\label{fig:synergyICA}
	\end{subfigure}
	\vskip\baselineskip
	\begin{subfigure}[b]{0.475\textwidth}   
		\centering 
		\includegraphics[width=\textwidth]{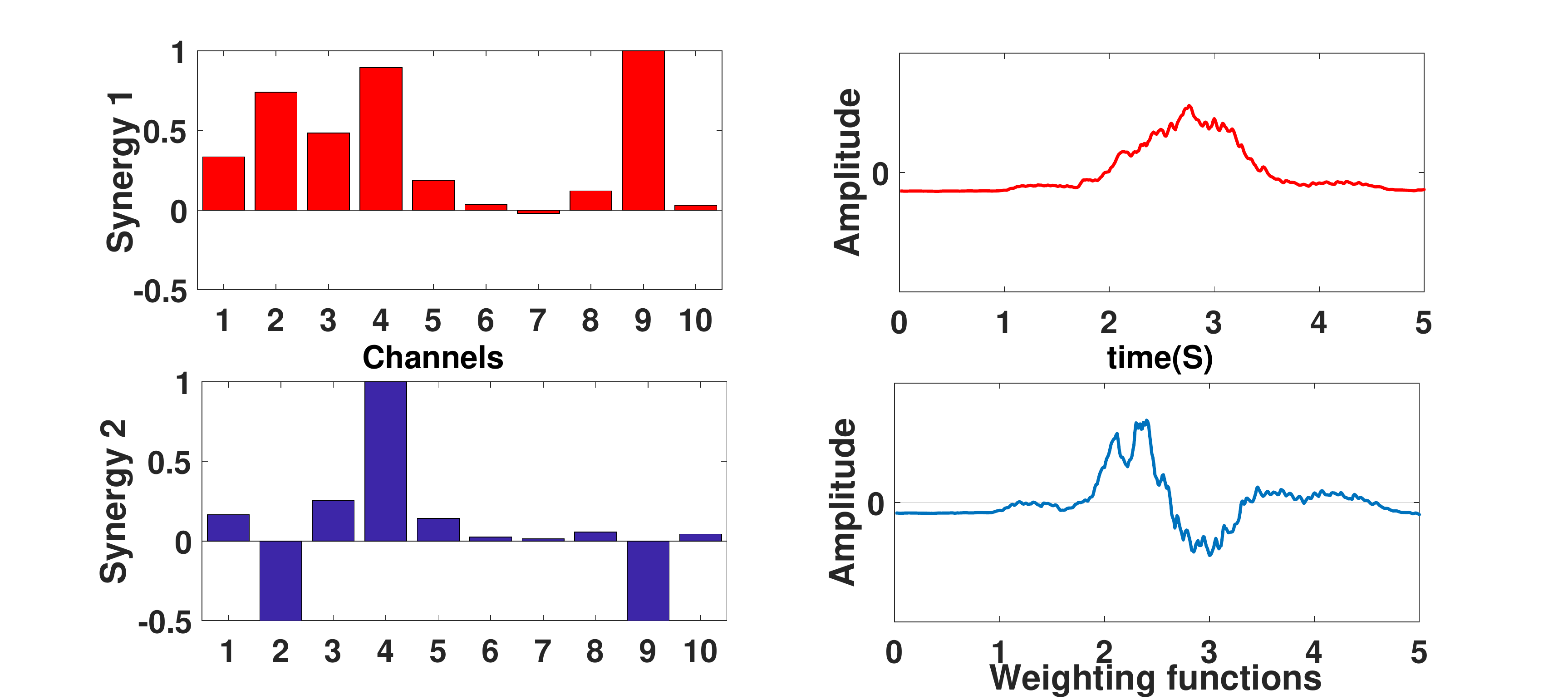}
		\caption[]%
		{{\small SOBI synergies}}    
		\label{fig:synergySOBI}
	\end{subfigure}
	\quad
	\begin{subfigure}[b]{0.475\textwidth}   
		\centering 
		\includegraphics[width=\textwidth]{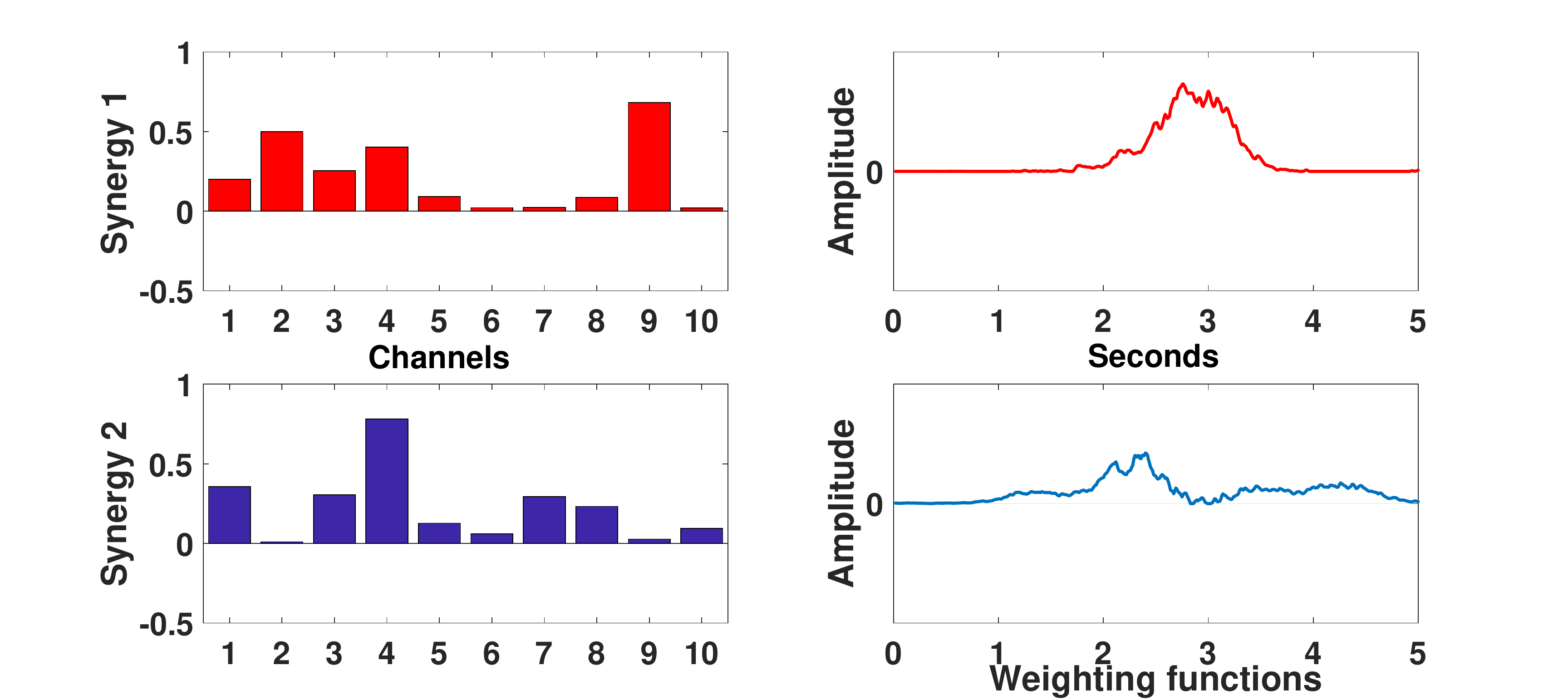}
		\caption[]%
		{{\small NMF synergies}}    
		\label{fig:synergyNMF}
	\end{subfigure}
	\caption{Two-component muscle synergy extracted via the four matrix factorisation methods for the 10-channel EMG signal recorded during wrist extension movement for 5 seconds (Subject 4/repetition 1)}
	
	\label{fig:Real_synergy_example}
\end{figure*}

As an illustration, the real 10-channel EMG epoch shown in Figure \ref{fig:subplotemg}  is factorised with the four matrix factorisation methods (PCA, ICA, SOBI and NMF) into two synergy model as shown in Figure \ref{fig:Real_synergy_example}.

\subsection{Factorisation performance comparison using synthetic data}

\begin{figure*}[]
	\centering
	\includegraphics[width=0.8\textwidth]{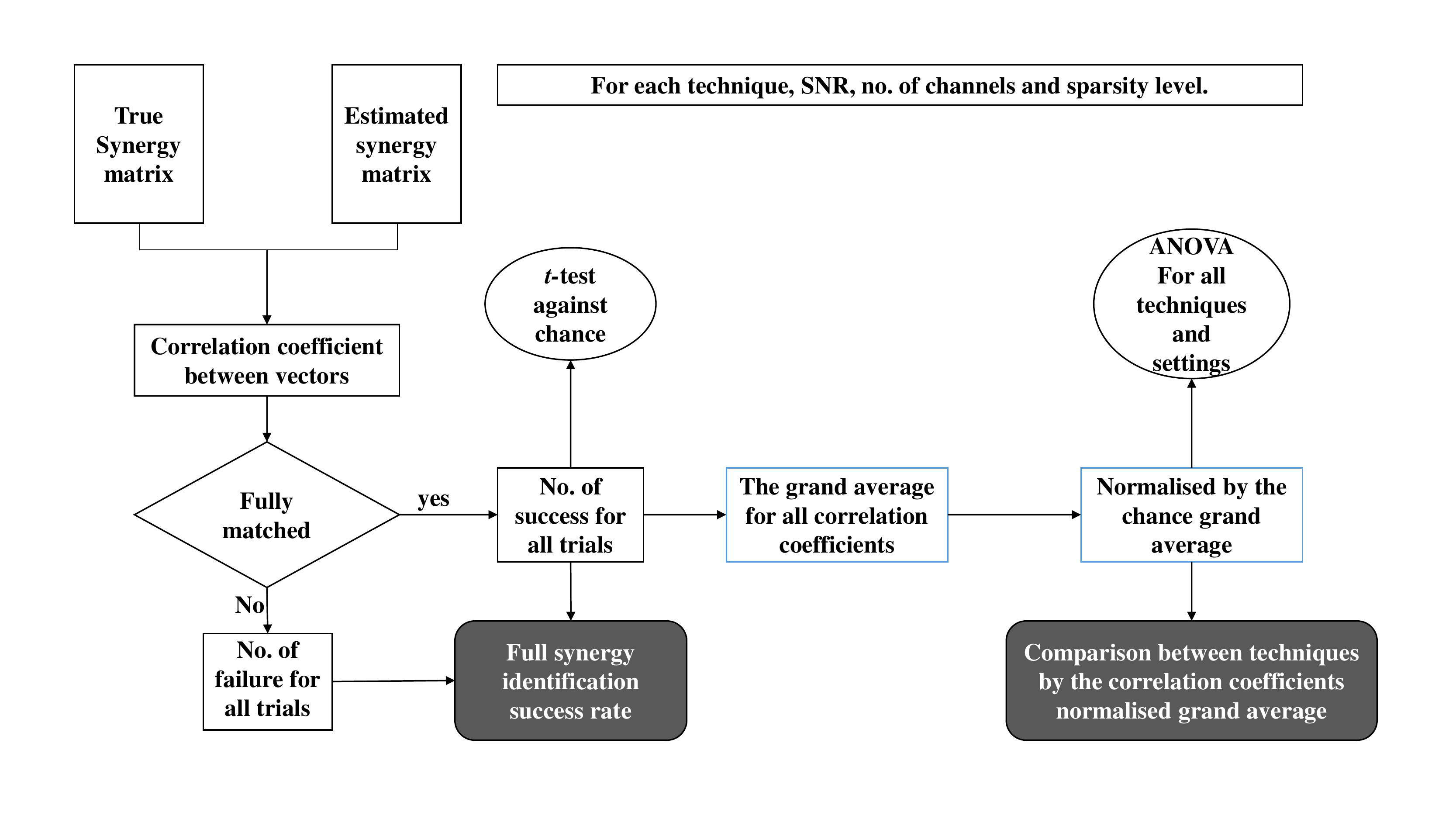}
	\caption{Block diagram for the comparison between matrix factorisation techniques.}
	\label{fig:blockdiagram}
\end{figure*}

The synthetic data was used to compare the ability of the four matrix factorisation techniques to estimate the muscle synergies in three different settings (SNR, number of channels and synergies sparsity). The comparison relies on the similarity between estimated and true synergies using the correlation coefficient on the basis of full identification of true synergies and similarity level between them. The sequence of this process is shown in Figure \ref{fig:blockdiagram}.

The first step is to match each of the extracted synergies with the true ones by calculating Pearson's correlation coefficients between them. True and estimated synergies with the highest correlation value are matched together. This matching is done freely and unconstrained. In other words, without forcing a full match (all four estimated synergies matched with all four true synergies) because in some cases two or more estimated synergies have the maximum correlation with the same true synergy. In those cases, the factorisation is not successful since the extracted synergies failed to fully represent all true synergies. Hence, the ``fully matched'' criterion is the ability of the factorisation method to estimate fully distinctive synergies that match all true synergies without duplication. The success rate for a ``fully matched'' is computed across the 10 generated datasets. It is used as a metric to judge the ability of extracted synergies to fully represent all the true synergies, since a good factorisation would represent all of them. 

In order to account to the chance that synergies may be randomly paired, the correlation coefficients between the true synergies and a set of randomly generated synergies are computed and the pairing rates are compared against for each factorisation method using a two-sample \textit{t}-test with significance level set up at (\textit{p} $<$ 0.05).

Secondly, the correlation coefficient values for fully identified synergies are averaged for each trial. The grand average is computed  for 10000 trials (1000 epochs$\times$ 10 datasets) of each setting combination. Then, it is normalised by the random synergy's correlation coefficients (chance grand average) as baseline removal as the following: $$Normalised\:grand\:average=\frac{ ( grand\:average - chance\:grand\:average)}{(1- chance\:grand\:average)}$$. The normalised grand average of the correlation coefficients between estimated and true synergies is computed for each matrix factorisation method with all different combination of the 3 settings (SNR levels, number of channels and sparsity). This criterion is an indicator of general factorisation quality. Therefore, we statistically analysed it to compare the factorisation techniques and the effect of all 3 settings using the 2-way ANOVA method with the significance level at (\textit{p} $<$ 0.05).

\subsection{Factorisation performance comparison using Real data}

Since there is no ground truth to compare each technique with for the real data, we compared the techniques regarding their application for prosthesis control. In several studies \cite{Ma2015a,Jiang2012a}, muscle synergy is used as a feature to classify different hand and wrist movements. Therefore, the factorisation techniques are assessed according to their classification accuracy for the 3 main wrists DoF.

To this end, the Ninapro real dataset is divided into training and testing sets with 60\% (6 repetitions of each task) of the data assigned to training for each subject. For each factorisation technique, synergies are estimated from training repetitions for each task. Those synergies are used to train \textit{k}-nearest neighbours (\textit{k}-NN) classifier (\textit{k}=3 for simplicity). Four classifiers are trained using the training synergies, three of them to classify between 2 tasks of each wrist DoF while the $4^{th}$ classifier is trained to classify between all 6 tasks. The number of synergies extracted was one for each repetition (two for each DoF) as in \cite{Jiang2014b} to avoid permutation issues. The testing dataset - which contains the other four repetitions of each task - is used to test those classifiers. One synergy is estimated directly from each task repetition in the test set using the four factorisation methods and used to predict the task through the trained classifiers. The classification error count for each DoF is used to evaluate the factorisation techniques.

\subsection{Number of synergies}

For the classification accuracy comparison using real datasets, the functional approach to determine number of synergies were chosen. A one-synergy model was applied for EMG activity of each movement. On the other hand, for the synthetic dataset comparison, the number of underlying synergies was known to be four.

The generated synthetic dataset can also be used to test the mathematical methods to determine the number of synergies. The minimum description length (MDL) was chosen as an alternative to  the explained variance methods as the latter is biased towards PCA since this relies on maximising the explained variance on the first components. The MDL method determine the number of synergies that could minimise the MDL. For more details please see Appendix \ref{appendix:MDL}.

In this study we use the synthetic dataset to test the ability of  MDL method to estimate the required number of synergies across various settings (Sparsity, noise and channel to synergy ratio). Since four true synergies are used, only the 8 and 12 channels datasets were investigated as the MDL boundary cannot estimate number of synergies when it is equal to channels. This is not a problem in practical applications since the muscle synergy hypothesis implies the concept of dimension reduction. In addition, three level of SNR (10, 15 and 20 dB) of sparse and non-sparse datasets were explored with 1000 trials for each combination. The result for correct estimation of synergies number is analysed via analysis of variance (ANOVA) and multiple comparison of population.

\section{Results}

\subsection{Number of synergies}

The model selection method based on MDL was examined with the synthetic EMG data where the number of synergies are known (four synergies). The MDL method was tested on 1000 trials for each combination of  sparsity, three levels of noise and two number of channels (8 and 12 channels).

\begin{figure}[]
	\centering
	\includegraphics[width=1\linewidth]{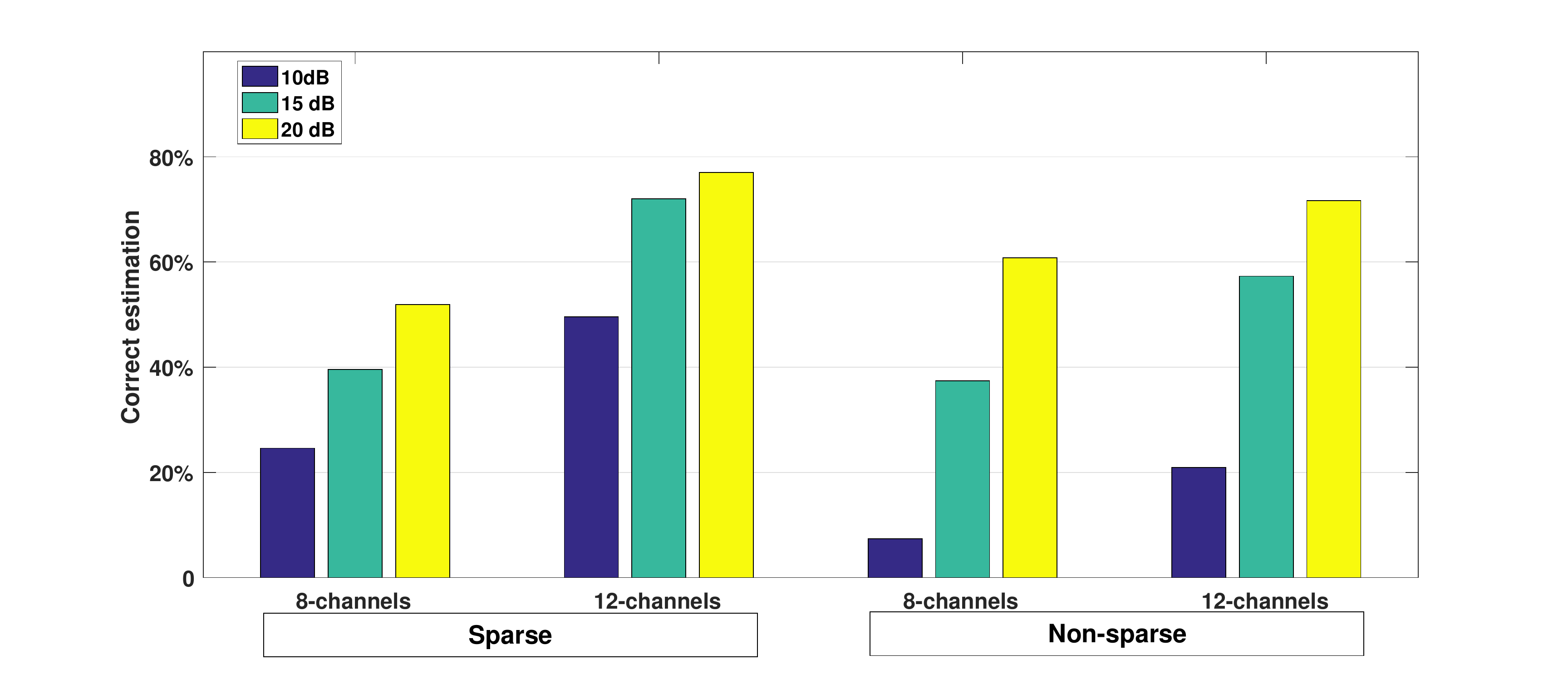}
	\caption{Percentage of correct synergy number estimation using the MDL method across the three settings (noise, number of channels and sparsity).}
	\label{fig:realdatamcp}
\end{figure}

The ANOVA shows that sparsity has no significant effect on the estimation of the correct number of synergies \textit{p}$>$0.05, while number of channels has a significant effect with $[F(1,11)=19.94, p=0.003]$ as 12-channels datasets performs better than 8-channel signals (shown in Figure \ref{fig:realdatamcp}). As for the level of noise, the 10 dB SNR had a significantly worse performance than 15 and 20 dB SNR with the effect of noise significant at $[F(2,11)=24.22, p=0.007]$ by 1-way ANOVA. This indicates that, the MDL method for estimating the correct number of synergies performs better with lower noise and more available channels, as expected. 

\subsection{Factorisation performance comparison using synthetic data}

\begin{figure}[]
	\centering
	
	\begin{subfigure}{0.5\textwidth}
		\centering
		\includegraphics[width=\textwidth]{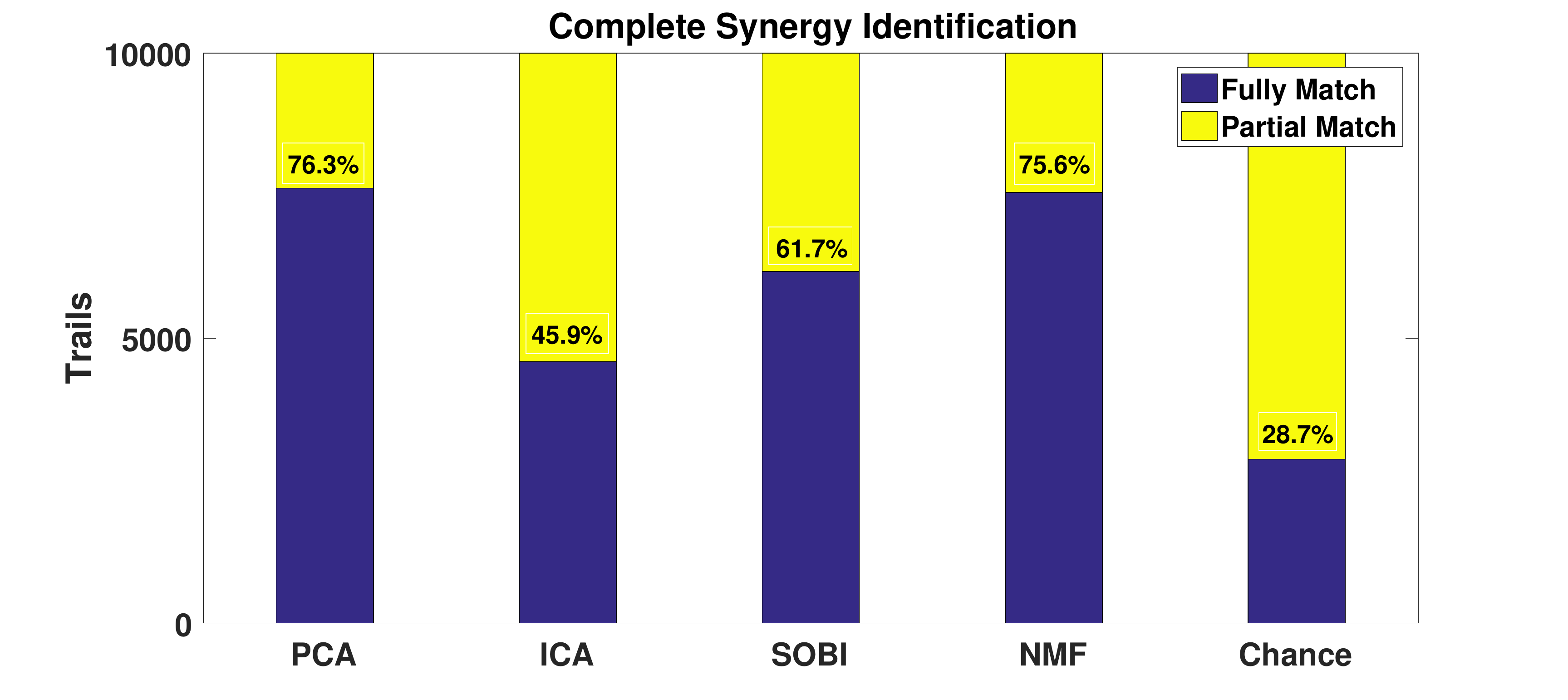}
		\caption{}
		\label{fig:Results}
	\end{subfigure}
	
	\hfill
	
	\begin{subfigure}{0.5\textwidth}
		\centering
		\includegraphics[width=\textwidth]{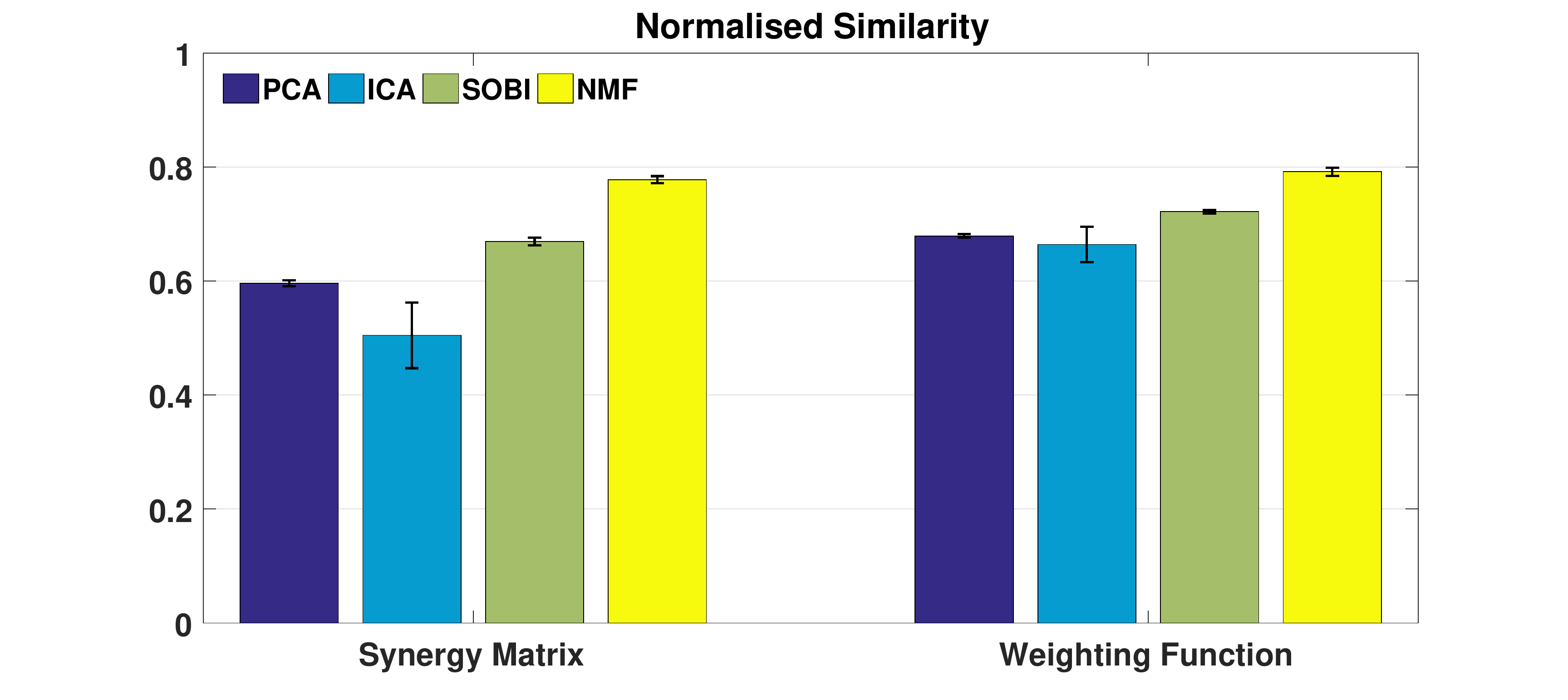}
		\caption{}
		\label{fig:Stats}
	\end{subfigure}
	
	\caption{The results for non-sparse, 12 channels dataset with 15dB SNR. Panel \ref{fig:Results}, the success ratio for the factorisation techniques to fully match the true synergies is shown. Panel \ref{fig:Stats}, the normalised similarity values for each technique single trial with the same settings. Error bars indicate standard deviation.}
	
	\label{fig:Results&Stats}
\end{figure}

The four matrix factorisation methods were compared on the basis of two criteria: synergy full identification success rate and the normalised grand average of correlation coefficients for the fully identified synergies. The comparison was done on 10000 trials (10 datasets of 1000 trails) for each combination of the three settings (sparsity, SNR and number of channels). An example of one setting of non-sparse, 12-channel with 15 dB SNR is shown in Figure \ref{fig:Results&Stats}. All the four factorisation techniques had converged for all trails except for ICA which failed to converge in 1.48\% of the trails.

\begin{figure}[]
	\centering
	\includegraphics[width=1\linewidth]{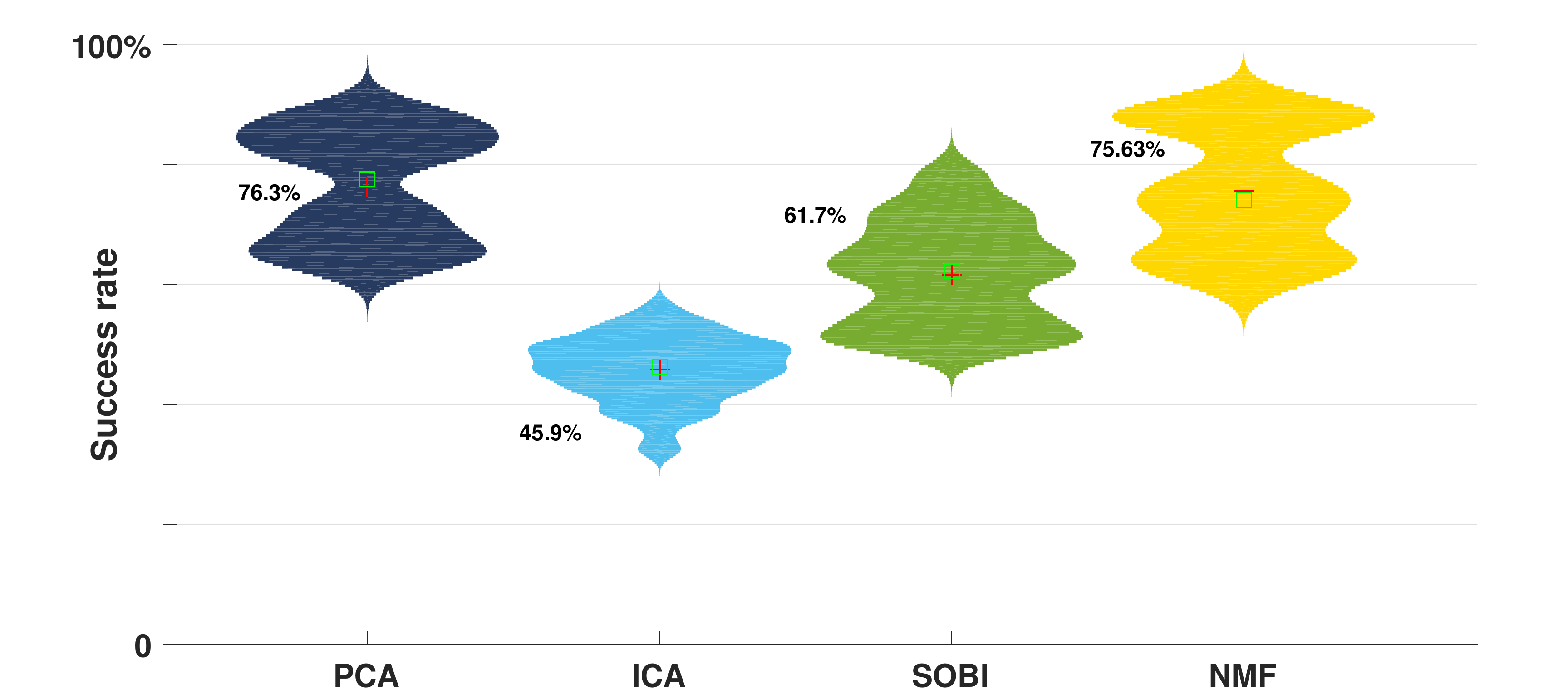}
	\caption{Violin graph for the success rate of full synergy identification for each method across all settings. The mean and median are represented in the Figure as red crosses and green squares respectively.}
	\label{fig:totalsucessrate}
\end{figure}

The four factorisation methods were assessed by their ability to fully identify all 4 true synergies by matching them according to their Pearson's correlation coefficients values. In order to rule out any statistical chance from it, a two-sample \textit{t}-test was conducted to compare the success rate of each technique and the randomly generated synergies. All the techniques succeeded to reject the null hypothesis  (\textit{p} $<$ 0.05) for all the settings. Hence, there is a significant difference between the matching success rate for each of the matrix factorisation methods and the randomly generated synergies. An example of the success rate for one of the settings is shown in Figure \ref{fig:Results}, while the average success rate to fully identify the true synergies for all settings is represented in Figure \ref{fig:totalsucessrate}. NMF and PCA are has the highest success rates to fully identify synergies.

\begin{figure*}[]
	\centering
	\includegraphics[width=0.9\textwidth]{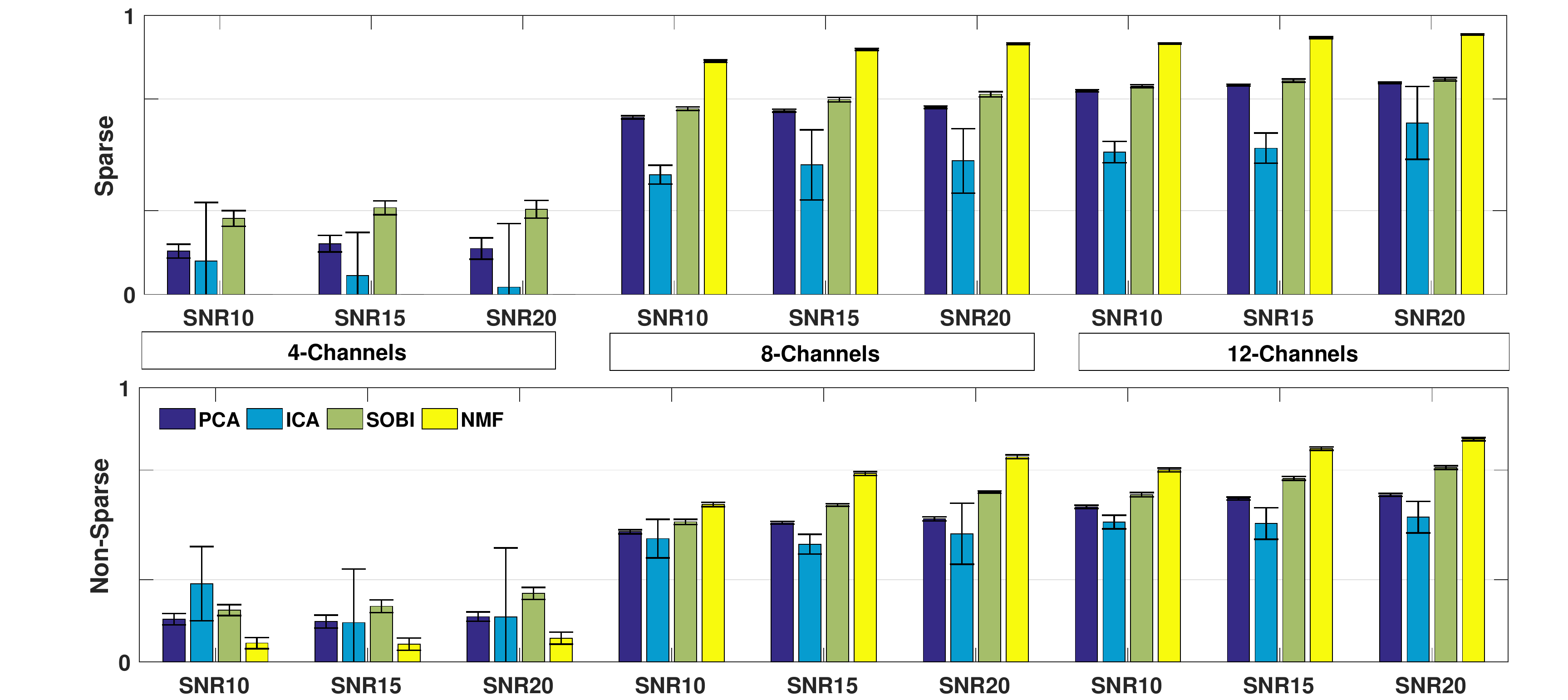}
	\caption{The normalised grand average of correlation coefficients for the fully identified synergies compared across all 3 settings (sparsity, SNR and number of channels) for the 4 matrix factorisation methods.  Error bars indicate standard deviation.}
	\label{fig:sumnormavg}
\end{figure*}

The correlation coefficients of the matched synergies were normalised by the random synergy correlation coefficients as shown in Figure \ref{fig:Stats}. Then the normalised correlation coefficient of synergies (synergy matrix) were averaged across trials. The grand average for each factorisation method was normalised by the chance's grand average. In Figure \ref{fig:sumnormavg}, the normalised grand average (similarity metric) for the four matrix factorisation methods is plotted for all different settings (sparsity, number of channels and noise level). It is worth mentioning that although NMF have the highest similarity for all settings except for the four channel case (the results for the sparse, four-channel setting for NMF are mostly negative). On the other hand, all four algorithms perform worse with four channels (no dimension reduction) with SOBI being the best algorithm among them in this case.

In order to explore the significance of those settings the two-way ANOVA was performed with post-hoc multiple comparison test. The result shows that number of channels and sparsity had a significant effect on the grand normalised average at [F(2,688)=1364.5, \textit{p} $\le$ 0.05] and [F(1,400)=7.35, \textit{p}=0.007] respectively. The multiple comparison test shows that sparse synergies and the higher number of channels show better similarity levels. On the other hand, the noise level fails to reject the null hypothesis. This means that the level of noise used in these experiments did not affect the quality of estimated synergies significantly unlike the sparsity or number of channels. In addition, this was supported by the interaction results, where factorisation methods and number of channels interaction showed a significant effect on the grand normalised average, as well as factorisation method and sparsity interaction. On the contrary, the noise level and factorisation techniques interaction have no significance on the grand normalised average.

The computational efficiency was compared after each technique ran for 100 times on Matlab 9 with Intel core i7 processor(2.4 GHz, 12 GB RAM) and the median value for the running time were computed. PCA and SOBI were the fastest with (0.0012 \textit{s} and 0.0015 \textit{s}) respectively followed by NMF with 0.0063 \textit{s} while ICA was significantly slower by 0.6419 \textit{s}.

\subsection{Factorisation performance comparison using Real data}

An example of the four matrix factorisation methods is shown in Figure \ref{fig:Real_synergy_example} by applying them on 10-channel EMG data. In order to show the similarities and differences in the estimated synergies and their weightings functions of each technique. For example, synergies extracted by PCA and SOBI have similarities in this example since both techniques are based on covariance matrices. The number of synergies needed in this example was chosen to be two according to the MDL method.

In addition, to compare between the matrix factorisation techniques, a one-component synergy was used to train a \textit{k}-NN classifier (\textit{k}=3) in order to classify between two antagonistic movements (one DoF) for each technique. This was calculated for the three wrist DoFs separately as shown in Table \ref{table_real_results}. In addition, the same synergies were used to classify between all six movements (three DoFs). The average classification error rate and its standard deviation for the 27 subjects is also represented in Table \ref{table_real_results}.

\begin{table}[]
	\centering
	\caption{The classification error count  and (error percentage) for each wrist's DoF (Sample size=216) and all 3 DoFs (sample size=648) across 27 subjects}
	\label{table_real_results}
	\begin{tabular}{@{}ccccc@{}}
		\toprule
		& \textbf{PCA}                                                   & \textbf{ICA}                                                    & \textbf{SOBI}                                                  & \textbf{NMF}                                                  \\ \midrule
		\textbf{\begin{tabular}[c]{@{}c@{}}DoF1\\ (wrist flexion\\ and extension)\end{tabular}}    & \begin{tabular}[c]{@{}c@{}}1\\ (0.46\%)\end{tabular}     & 		    \begin{tabular}[c]{@{}c@{}}28\\ (12.96\%)\end{tabular}      &  \begin{tabular}[c]{@{}c@{}}8\\ (3.70\%)\end{tabular}       & 
		\begin{tabular}[c]{@{}c@{}}1\\ (0.46\%)\end{tabular}   \\
		\rowcolor[HTML]{EFEFEF} 
		\textbf{\begin{tabular}[c]{@{}c@{}}DoF2\\ (wrist radial and\\ ulnar deviation)\end{tabular}}    & \begin{tabular}[c]{@{}c@{}}12\\ (5.56\%)\end{tabular}     & \begin{tabular}[c]{@{}c@{}}29\\ (13.43\%)\end{tabular}      & \begin{tabular}[c]{@{}c@{}}19\\ (8.80\%)\end{tabular}      & \begin{tabular}[c]{@{}c@{}}1\\ (0.46\%)\end{tabular}   \\
		\textbf{\begin{tabular}[c]{@{}c@{}}DoF3\\ (wrist supination\\ and pronation)\end{tabular}}    & \begin{tabular}[c]{@{}c@{}}7\\ (3.24\%)\end{tabular}     & \begin{tabular}[c]{@{}c@{}}31\\ (14.35\%)\end{tabular}    & \begin{tabular}[c]{@{}c@{}}18\\ (8.33\%)\end{tabular}     & \begin{tabular}[c]{@{}c@{}}5\\ (2.31\%)\end{tabular}  \\
		\rowcolor[HTML]{EFEFEF} 
		\textbf{\begin{tabular}[c]{@{}c@{}}All 3 DoFs\\ (all 6 movements)\end{tabular}}              & \begin{tabular}[c]{@{}c@{}} 43\\ (6.64\%)\end{tabular} & \begin{tabular}[c]{@{}c@{}}122\\  (18.83\%)\end{tabular} & \begin{tabular}[c]{@{}c@{}}65\\ (10.03\%)\end{tabular} & \begin{tabular}[c]{@{}c@{}}41\\ (6.33\%)\end{tabular} \\ \bottomrule
	\end{tabular}
\end{table}

\section{Discussion and Conclusion}

In this paper, we compared the most common matrix factorisation techniques (PCA, ICA and NMF) for muscle synergy estimation alongside SOBI, a BSS method that had not been applied for synergy extraction yet. Many studies rely on muscle synergy concept such as myoelectric control and biomechanical research. However, only two studies \cite{Tresch2006,Lambert-Shirzad2017} compared various factorisation methods (excluding SOBI) for synergy estimation without investigating the factors that affect the factorisation quality - except for noise. 
  
Herein, the comparison was held on real data and synthetic signals generated with known synergies and under different settings. Using the synthetic data we studied the effect of those settings on the muscle synergy extraction for each technique. The sparsity nature of synergies and level of noise was investigated in addition to the number of channels needed to extract the four synthetic synergies. The ability of the four factorisation methods to extract  synergies from  synthetic data was judged according to two metrics: success rate  to fully identify synergies (Figure \ref{fig:totalsucessrate}) and the correlation coefficients between true and estimated synergies (Figure \ref{fig:sumnormavg}). Moreover, the synthetic data was used to assess the MDL method to determine number of synergies needed under those three settings.

For the real datasets, since there is no ground truth to compare synergies estimated, we compared the factorisation methods according to the ability of their extracted synergies to classify wrist movements (Table \ref{table_real_results}) as a proof of concept for prosthesis control \cite{Rasool2016,Choi2011}. PCA and NMF had the best classification accuracy followed by SOBI, while ICA had the lowest accuracy.

On the other hand,  the synthetic datasets results showed that NMF and PCA had better success rate to fully identify the four true synergies than SOBI and ICA. However, NMF and SOBI had the best normalised grand average of correlation coefficients (similarity level) between estimated and true synergies followed by PCA then ICA. Notably,  NMF performed poorly with four-channel datasets when there was not any dimension reduction. In general, all algorithms perform better with higher number of channels compared to synergies, where SOBI was the best algorithm when there is no dimension reduction. Therefore, SOBI would be a  relevant algorithm in situations  with limited number of electrodes as it is preferable to minimise the number of electrodes for practical prosthesis control \cite{Clancy2017,Muceli2014}.

The two-way ANOVA showed that the tested range of SNR has no significance effect on the factorisation performance, although it is noticed that ICA was the most unaffected method to noise according to the multiple comparison test. On the other hand, sparsity had a significant effect (\textit{p}$<$ 0.05) on the correlation between true and estimated synergies. According to the multiple comparison test, the sparse synergies are easier to estimate by all factorisation methods. Moreover, number of channels shows a significant effect (\textit{p}$<$ 0.05) on the correlation between estimated synergies and true ones. In addition, higher number of channels to number of synergies ratio provides better synergy extraction. 


Regarding the estimation of the number of synergies, the multichannel EMG signal is reduced into a lower subspace for the purpose of synergy extraction. The estimation of this subspace's dimension or, in other words, the number of synergies is crucial for the factorisation process. In the literature, there are two main approaches to determine the appropriate number of synergies: the functional and the mathematical ones. The functional approach determines the number of synergies according to the myoelectric control requirements such as assigning two \cite{Muceli2010d,Jiang2009} synergies for each DoF. On the other hand, the mathematical approach relies on explained variance (using tests such as  scree plot and Bart test) or the likelihood criteria (such as Akaike information criteria and MDL) \cite{Ikeda2000}. Here, we explored the MDL as an alternative for variance explained methods. The  results show that MDL performs better with higher channel to synergy ratio. This supports the current challenges for effective synergy identification with limited number of electrodes. However, further investigation is needed to compare between different number of synergies estimation methods using synthetic datasets with various settings. 

Other limitations are worth noting. The results may be biased towards NMF due to the non-negative nature of the simulated synergies. However, this choice is supported by previous studies \cite{Choi2011} which suggested the usefulness of NMF due to the additive nature of the synergies. In addition, further examination is needed if the setting of EMG acquisition changes dramatically (really bad SNR, much higher number of channels, etc.) to evaluate the validity of our conclusions in those settings. Finally, since various studies employ the muscle synergy in prosthesis control, a simple approach (\textit{k}-NN classifier) was used in this paper as an example to guide synergy application and to support the synthetic results.  We treated this part of the study as a proof of concept. Additional work is needed with more advanced techniques and variety of tasks and movements. 

In conclusion, this paper compared matrix factorisation algorithms for muscle synergy extraction and the factors that affect the quality of estimated synergies. Our findings suggest that the presence of sparse synergies and higher number of channels would improve the quality of extracted synergies. When the number of channels equal to synergies (no dimension reduction), SOBI performed better than other methods although the performance was still poor in this case. Otherwise, NMF is the best solution for robust synergy extraction when number of channels/muscles is higher than the required muscle synergies. 

\section*{Declarations}

 \noindent Competing interests: None declared \\ 
 Funding: None\\
 Ethical approval: Not required
 

\section*{References}
\bibliography{MatrixFactSynergyMendeley}

\begin{thebibliography}{10}
\expandafter\ifx\csname url\endcsname\relax
  \def\url#1{\texttt{#1}}\fi
\expandafter\ifx\csname urlprefix\endcsname\relax\def\urlprefix{URL }\fi
\expandafter\ifx\csname href\endcsname\relax
  \def\href#1#2{#2} \def\path#1{#1}\fi

\bibitem{DAvella2015}
A.~D'Avella, M.~Giese, Y.~P. Ivanenko, T.~Schack, T.~Flash, {Editorial:
  Modularity in motor control: from muscle synergies to cognitive action
  representation.}, Frontiers in computational neuroscience 9 (2015) 126.
\newblock \href {http://dx.doi.org/10.3389/fncom.2015.00126}
  {\path{doi:10.3389/fncom.2015.00126}}.

\bibitem{Sherrington1910a}
C.~S. Sherrington, {Flexion-reflex of the limb, crossed extension-reflex, and
  reflex stepping and standing}, The Journal of Physiology 40~(1-2) (1910)
  28--121.
\newblock \href {http://dx.doi.org/10.1113/jphysiol.1910.sp001362}
  {\path{doi:10.1113/jphysiol.1910.sp001362}}.

\bibitem{Bizzi1991}
E.~Bizzi, F.~A. Mussa-Ivaldi, S.~F. Giszter, {Computations underlying the
  execution of movement: a biological perspective.}, Science (New York, N.Y.)
  253~(5017) (1991) 287--91.
\newblock \href {http://dx.doi.org/10.1126/science.1857964}
  {\path{doi:10.1126/science.1857964}}.

\bibitem{Giszter1993}
S.~F.~S. Giszter, F.~F.~A. Mussa-Ivaldi, E.~Bizzi, {Convergent force fields
  organized in the frog's spinal cord.}, The Journal of Neuroscience 13~(2)
  (1993) 467--491.

\bibitem{Mussa-Ivaldi1994}
F.~A. Mussa-Ivaldi, S.~F. Giszter, E.~Bizzi, {Linear combinations of primitives
  in vertebrate motor control.}, Proceedings of the National Academy of
  Sciences of the United States of America 91~(16) (1994) 7534--7538.
\newblock \href {http://dx.doi.org/10.1073/pnas.91.16.7534}
  {\path{doi:10.1073/pnas.91.16.7534}}.

\bibitem{Tresch1999}
M.~C. Tresch, P.~Saltiel, E.~Bizzi, {The construction of movement by the spinal
  cord.}, Nature neuroscience 2~(2) (1999) 162--7.
\newblock \href {http://dx.doi.org/10.1038/5721} {\path{doi:10.1038/5721}}.

\bibitem{Saltiel2001}
P.~Saltiel, K.~Wyler-Duda, A.~D'Avella, M.~C. Tresch, E.~Bizzi, {Muscle
  synergies encoded within the spinal cord: evidence from focal intraspinal
  NMDA iontophoresis in the frog.}, Journal of neurophysiology 85~(2) (2001)
  605--619.

\bibitem{DAvella2003}
A.~D'Avella, P.~Saltiel, E.~Bizzi, {Combinations of muscle synergies in the
  construction of a natural motor behavior.}, Nature neuroscience 6~(3) (2003)
  300--308.
\newblock \href {http://dx.doi.org/10.1038/nn1010} {\path{doi:10.1038/nn1010}}.

\bibitem{TRESCH}
M.~C. Tresch, A.~Jarc, {The case for and against muscle synergies.}, Current
  opinion in neurobiology 19~(6) (2009) 601--7.
\newblock \href {http://dx.doi.org/10.1016/j.conb.2009.09.002}
  {\path{doi:10.1016/j.conb.2009.09.002}}.

\bibitem{Kargo2008}
W.~J. Kargo, S.~F. Giszter, {Individual Premotor Drive Pulses, Not Time-Varying
  Synergies, Are the Units of Adjustment for Limb Trajectories Constructed in
  Spinal Cord}, Journal of Neuroscience 28~(10) (2008) 2409--2425.
\newblock \href {http://dx.doi.org/10.1523/JNEUROSCI.3229-07.2008}
  {\path{doi:10.1523/JNEUROSCI.3229-07.2008}}.

\bibitem{Hart2004}
C.~B. Hart, S.~F. Giszter, {Modular premotor drives and unit bursts as
  primitives for frog motor behaviors.}, The Journal of neuroscience : the
  official journal of the Society for Neuroscience 24~(22) (2004) 5269--82.
\newblock \href {http://dx.doi.org/10.1523/JNEUROSCI.5626-03.2004}
  {\path{doi:10.1523/JNEUROSCI.5626-03.2004}}.

\bibitem{Tresch1999a}
M.~C. Tresch, E.~Bizzi, {Responses to spinal microstimulation in the
  chronically spinalized rat and their relationship to spinal systems activated
  by low threshold cutaneous stimulation}, Experimental Brain Research 129~(3)
  (1999) 0401--0416.
\newblock \href {http://dx.doi.org/10.1007/s002210050908}
  {\path{doi:10.1007/s002210050908}}.

\bibitem{Lemay2004}
M.~A. Lemay, W.~M. Grill, {Modularity of motor output evoked by intraspinal
  microstimulation in cats.}, Journal of neurophysiology 91~(1) (2004) 502--14.
\newblock \href {http://dx.doi.org/10.1152/jn.00235.2003}
  {\path{doi:10.1152/jn.00235.2003}}.

\bibitem{Haiss2005}
F.~Haiss, C.~Schwarz, {Spatial segregation of different modes of movement
  control in the whisker representation of rat primary motor cortex.}, The
  Journal of neuroscience : the official journal of the Society for
  Neuroscience 25~(6) (2005) 1579--87.
\newblock \href {http://dx.doi.org/10.1523/JNEUROSCI.3760-04.2005}
  {\path{doi:10.1523/JNEUROSCI.3760-04.2005}}.

\bibitem{Stepniewska2005}
I.~Stepniewska, P.-C. Fang, J.~H. Kaas, {Microstimulation reveals specialized
  subregions for different complex movements in posterior parietal cortex of
  prosimian galagos.}, Proceedings of the National Academy of Sciences of the
  United States of America 102~(13) (2005) 4878--83.
\newblock \href {http://dx.doi.org/10.1073/pnas.0501048102}
  {\path{doi:10.1073/pnas.0501048102}}.

\bibitem{Overduin2008}
S.~A. Overduin, A.~D'Avella, J.~Roh, E.~Bizzi, {Modulation of Muscle Synergy
  Recruitment in Primate Grasping}, Journal of Neuroscience 28~(4) (2008)
  880--892.
\newblock \href {http://dx.doi.org/10.1523/JNEUROSCI.2869-07.2008}
  {\path{doi:10.1523/JNEUROSCI.2869-07.2008}}.

\bibitem{Overduin2014a}
S.~a. Overduin, A.~D'Avella, J.~M. Carmena, E.~Bizzi, {Muscle synergies evoked
  by microstimulation are preferentially encoded during behavior.}, Frontiers
  in computational neuroscience 8~(March) (2014) 20.
\newblock \href {http://dx.doi.org/10.3389/fncom.2014.00020}
  {\path{doi:10.3389/fncom.2014.00020}}.

\bibitem{Ting2005}
L.~H. Ting, J.~M. Macpherson, {A limited set of muscle synergies for force
  control during a postural task.}, Journal of neurophysiology 93~(1) (2005)
  609--13.
\newblock \href {http://dx.doi.org/10.1152/jn.00681.2004}
  {\path{doi:10.1152/jn.00681.2004}}.

\bibitem{Torres-Oviedo2006}
G.~Torres-Oviedo, J.~M. Macpherson, L.~H. Ting, {Muscle synergy organization is
  robust across a variety of postural perturbations.}, Journal of
  neurophysiology 96~(3) (2006) 1530--1546.
\newblock \href {http://dx.doi.org/10.1152/jn.00810.2005}
  {\path{doi:10.1152/jn.00810.2005}}.

\bibitem{Cheung2005}
V.~C.-K.~K. Cheung, {Central and Sensory Contributions to the Activation and
  Organization of Muscle Synergies during Natural Motor Behaviors}, Journal of
  Neuroscience 25~(27) (2005) 6419--6434.
\newblock \href {http://dx.doi.org/10.1523/JNEUROSCI.4904-04.2005}
  {\path{doi:10.1523/JNEUROSCI.4904-04.2005}}.

\bibitem{Weiss2004}
E.~J. Weiss, M.~Flanders, {Muscular and postural synergies of the human hand.},
  Journal of neurophysiology 92~(1) (2004) 523--35.
\newblock \href {http://dx.doi.org/10.1152/jn.01265.2003}
  {\path{doi:10.1152/jn.01265.2003}}.

\bibitem{DAvella2006}
A.~D'Avella, A.~Portone, L.~Fernandez, F.~Lacquaniti, {Control of Fast-Reaching
  Movements by Muscle Synergy Combinations}, Journal of Neuroscience 26~(30)
  (2006) 7791--7810.
\newblock \href {http://dx.doi.org/10.1523/JNEUROSCI.0830-06.2006}
  {\path{doi:10.1523/JNEUROSCI.0830-06.2006}}.

\bibitem{Wakeling2009}
J.~M. J. J.~M. Wakeling, T.~Horn, {Neuromechanics of muscle synergies during
  cycling}, Journal of neurophysiology 101~(2) (2009) 843--54.
\newblock \href {http://dx.doi.org/10.1152/jn.90679.2008}
  {\path{doi:10.1152/jn.90679.2008}}.

\bibitem{Hug2011a}
F.~Hug, N.~a. Turpin, A.~Couturier, S.~Dorel, {Consistency of muscle synergies
  during pedaling across different mechanical constraints.}, Journal of
  neurophysiology 106~(1) (2011) 91--103.
\newblock \href {http://dx.doi.org/10.1152/jn.01096.2010}
  {\path{doi:10.1152/jn.01096.2010}}.

\bibitem{Ting2007}
L.~H. Ting, J.~L. McKay, {Neuromechanics of muscle synergies for posture and
  movement.}, Current opinion in neurobiology 17~(6) (2007) 622--8.
\newblock \href {http://dx.doi.org/10.1016/j.conb.2008.01.002}
  {\path{doi:10.1016/j.conb.2008.01.002}}.

\bibitem{Aoi2015}
S.~Aoi, T.~Funato, {Neuromusculoskeletal models based on the muscle synergy
  hypothesis for the investigation of adaptive motor control in locomotion via
  sensory-motor coordination}, Neuroscience Research 104 (2016) 88--95.
\newblock \href {http://dx.doi.org/10.1016/j.neures.2015.11.005}
  {\path{doi:10.1016/j.neures.2015.11.005}}.

\bibitem{Pons2016a}
D.~Torricelli, F.~Barroso, M.~Coscia, C.~Alessandro, F.~Lunardini, E.~{Bravo
  Esteban}, A.~D'Avella, {Muscle Synergies in Clinical Practice: Theoretical
  and Practical Implications}, in: J.~L. Pons, R.~Raya, J.~Gonz{\'{a}}lez
  (Eds.), Emerging Therapies in Neurorehabilitation II, Vol.~10 of Biosystems
  {\&} Biorobotics, Springer International Publishing, Cham, 2016, pp.
  251--272.
\newblock \href {http://dx.doi.org/10.1007/978-3-319-24901-8}
  {\path{doi:10.1007/978-3-319-24901-8}}.

\bibitem{Nazifi2017}
M.~M. Nazifi, H.~U. Yoon, K.~Beschorner, P.~Hur, {Shared and Task-Specific
  Muscle Synergies during Normal Walking and Slipping}, Frontiers in Human
  Neuroscience 11~(February) (2017) 1--14.
\newblock \href {http://dx.doi.org/10.3389/fnhum.2017.00040}
  {\path{doi:10.3389/fnhum.2017.00040}}.

\bibitem{Martino2015}
G.~Martino, Y.~P. Ivanenko, A.~D'Avella, M.~Serrao, A.~Ranavolo, F.~Draicchio,
  G.~Cappellini, C.~Casali, F.~Lacquaniti, {Neuromuscular adjustments of gait
  associated with unstable conditions}, Journal of Neurophysiology 114~(2011)
  (2015) jn.00029.2015.
\newblock \href {http://dx.doi.org/10.1152/jn.00029.2015}
  {\path{doi:10.1152/jn.00029.2015}}.

\bibitem{Rasool2016}
G.~Rasool, K.~Iqbal, N.~Bouaynaya, G.~White, {Real-Time Task Discrimination for
  Myoelectric Control Employing Task-Specific Muscle Synergies.}, IEEE
  transactions on neural systems and rehabilitation engineering : a publication
  of the IEEE Engineering in Medicine and Biology Society 24~(1) (2016)
  98--108.
\newblock \href {http://dx.doi.org/10.1109/TNSRE.2015.2410176}
  {\path{doi:10.1109/TNSRE.2015.2410176}}.

\bibitem{Ma2015a}
J.~Ma, N.~V. Thakor, F.~Matsuno, {Hand and Wrist Movement Control of
  Myoelectric Prosthesis Based on Synergy}, IEEE Transactions on Human-Machine
  Systems 45~(1) (2015) 74--83.
\newblock \href {http://dx.doi.org/10.1109/THMS.2014.2358634}
  {\path{doi:10.1109/THMS.2014.2358634}}.

\bibitem{Jiang2014b}
N.~Jiang, H.~Rehbaum, I.~Vujaklija, B.~Graimann, D.~Farina, {Intuitive, online,
  simultaneous, and proportional myoelectric control over two
  degrees-of-freedom in upper limb amputees.}, IEEE transactions on neural
  systems and rehabilitation engineering : a publication of the IEEE
  Engineering in Medicine and Biology Society 22~(3) (2014) 501--10.
\newblock \href {http://dx.doi.org/10.1109/TNSRE.2013.2278411}
  {\path{doi:10.1109/TNSRE.2013.2278411}}.

\bibitem{DAvella2002}
A.~D'Avella, M.~M. C.~M. Tresch, {Modularity in the motor system: decomposition
  of muscle patterns as combinations of time-varying synergies}, Citeseer 1
  (2002) 141--148.
\newblock \href {http://dx.doi.org/10.1.1.19.8895} {\path{doi:10.1.1.19.8895}}.

\bibitem{Hart2013}
C.~B. Hart, S.~F. Giszter, {Distinguishing synchronous and time-varying
  synergies using point process interval statistics: motor primitives in frog
  and rat.}, Frontiers in computational neuroscience 7~(May) (2013) 52.
\newblock \href {http://dx.doi.org/10.3389/fncom.2013.00052}
  {\path{doi:10.3389/fncom.2013.00052}}.

\bibitem{Jackson1991}
J.~E. Jackson, {A User's Guide to Principal Components}, Wiley Series in
  Probability and Statistics, John Wiley {\&} Sons, Inc., Hoboken, NJ, USA,
  1991.
\newblock \href {http://dx.doi.org/10.1002/0471725331}
  {\path{doi:10.1002/0471725331}}.

\bibitem{Ranganathan2012}
R.~Ranganathan, C.~Krishnan, {Extracting synergies in gait: using EMG
  variability to evaluate control strategies.}, Journal of Neurophysiology
  108~(5) (2012) 1537--44.
\newblock \href {http://dx.doi.org/10.1152/jn.01112.2011}
  {\path{doi:10.1152/jn.01112.2011}}.

\bibitem{Hyvarinen2000}
A.~Hyv{\"{a}}rinen, E.~Oja, {Independent component analysis: algorithms and
  applications}, Neural Networks 13~(4-5) (2000) 411--430.
\newblock \href {http://dx.doi.org/10.1016/S0893-6080(00)00026-5}
  {\path{doi:10.1016/S0893-6080(00)00026-5}}.

\bibitem{Kargo2003}
W.~J. Kargo, D.~A. Nitz, {Early skill learning is expressed through selection
  and tuning of cortically represented muscle synergies.}, The Journal of
  Neuroscience 23~(35) (2003) 11255--69.
\newblock \href {http://dx.doi.org/10.1162/089892903322307384}
  {\path{doi:10.1162/089892903322307384}}.

\bibitem{Lee1999}
D.~D. Lee, H.~S. Seung, {Learning the parts of objects by non-negative matrix
  factorization.}, Nature 401~(6755) (1999) 788--91.
\newblock \href {http://dx.doi.org/10.1038/44565} {\path{doi:10.1038/44565}}.

\bibitem{Choi2011}
C.~Choi, J.~Kim, {Synergy matrices to estimate fluid wrist movements by surface
  electromyography}, Medical Engineering and Physics 33~(8) (2011) 916--923.
\newblock \href {http://dx.doi.org/10.1016/j.medengphy.2011.02.006}
  {\path{doi:10.1016/j.medengphy.2011.02.006}}.

\bibitem{Berger2014}
D.~J. Berger, A.~D'Avella, {Effective force control by muscle synergies.},
  Frontiers in computational neuroscience 8~(April) (2014) 46.
\newblock \href {http://dx.doi.org/10.3389/fncom.2014.00046}
  {\path{doi:10.3389/fncom.2014.00046}}.

\bibitem{Belouchrani1997}
A.~Belouchrani, K.~Abed-Meraim, J.-F.~J. Cardoso, E.~Moulines, {A blind source
  separation technique using second-order statistics}, IEEE Transactions on
  Signal Processing 45~(2) (1997) 434--444.
\newblock \href {http://dx.doi.org/10.1109/78.554307}
  {\path{doi:10.1109/78.554307}}.

\bibitem{Tresch2006}
M.~C. Tresch, V.~C.-K.~K. Cheung, A.~D'Avella, {Matrix factorization algorithms
  for the identification of muscle synergies: evaluation on simulated and
  experimental data sets.}, Journal of neurophysiology 95~(4) (2006)
  2199--2212.
\newblock \href {http://dx.doi.org/10.1152/jn.00222.2005}
  {\path{doi:10.1152/jn.00222.2005}}.

\bibitem{Lambert-Shirzad2017}
N.~Lambert-Shirzad, H.~F.~M. {Van der Loos}, {On identifying kinematic and
  muscle synergies: a comparison of matrix factorization methods using
  experimental data from the healthy population}, Journal of Neurophysiology
  117~(1) (2017) 290--302.
\newblock \href {http://dx.doi.org/10.1152/jn.00435.2016}
  {\path{doi:10.1152/jn.00435.2016}}.

\bibitem{Atzori2014}
M.~Atzori, A.~Gijsberts, C.~Castellini, B.~Caputo, A.-G.~M. Hager, S.~Elsig,
  G.~Giatsidis, F.~Bassetto, H.~M{\"{u}}ller, {Electromyography data for
  non-invasive naturally-controlled robotic hand prostheses.}, Scientific data
  1 (2014) 140053.
\newblock \href {http://dx.doi.org/10.1038/sdata.2014.53}
  {\path{doi:10.1038/sdata.2014.53}}.

\bibitem{Atzori2015a}
M.~Atzori, A.~Gijsberts, I.~Kuzborskij, S.~Elsig, A.~G. Hager, O.~Deriaz,
  C.~Castellini, H.~Muller, B.~Caputo, A.-G.~M. Hager, O.~Deriaz,
  C.~Castellini, H.~Muller, B.~Caputo, {Characterization of a benchmark
  database for myoelectric movement classification}, IEEE Transactions on
  Neural Systems and Rehabilitation Engineering 23~(1) (2015) 73--83.
\newblock \href {http://dx.doi.org/10.1109/TNSRE.2014.2328495}
  {\path{doi:10.1109/TNSRE.2014.2328495}}.

\bibitem{Farina2014c}
D.~Farina, N.~Jiang, H.~Rehbaum, A.~Holobar, B.~Graimann, H.~Dietl, O.~C.
  Aszmann, {The extraction of neural information from the surface EMG for the
  control of upper-limb prostheses: Emerging avenues and challenges}, IEEE
  Transactions on Neural Systems and Rehabilitation Engineering 22~(4) (2014)
  797--809.
\newblock \href {http://dx.doi.org/10.1109/TNSRE.2014.2305111}
  {\path{doi:10.1109/TNSRE.2014.2305111}}.

\bibitem{Torres-Oviedo2007}
G.~Torres-Oviedo, L.~H. Ting, {Muscle synergies characterizing human postural
  responses.}, Journal of neurophysiology 98~(4) (2007) 2144--56.
\newblock \href {http://dx.doi.org/10.1152/jn.01360.2006}
  {\path{doi:10.1152/jn.01360.2006}}.

\bibitem{Abdi2010}
H.~Abdi, L.~J. Williams, {Principal component analysis}, Wiley
  Interdisciplinary Reviews: Computational Statistics 2~(4) (2010) 433--459.
\newblock \href {http://arxiv.org/abs/arXiv:1011.1669v3}
  {\path{arXiv:arXiv:1011.1669v3}}, \href {http://dx.doi.org/10.1002/wics.101}
  {\path{doi:10.1002/wics.101}}.

\bibitem{Hyviirinen}
A.~Hyv{\"{a}}rinen, {A family of fixed-point algorithms for independent
  component analysis}, in: IEEE 5\textsuperscript{th} Int. Conf. on Acoustics,
  Speech and Signal Processing (ICASSP), 1997, pp. 3917--3920.
\newblock \href {http://dx.doi.org/10.1109/ICASSP.1997.604766}
  {\path{doi:10.1109/ICASSP.1997.604766}}.

\bibitem{Berry2007}
M.~W. Berry, M.~Browne, A.~N. Langville, V.~P. Pauca, R.~J. Plemmons,
  {Algorithms and applications for approximate nonnegative matrix
  factorization}, Computational Statistics and Data Analysis 52~(1) (2007)
  155--173.
\newblock \href {http://dx.doi.org/10.1016/j.csda.2006.11.006}
  {\path{doi:10.1016/j.csda.2006.11.006}}.

\bibitem{Tanaka}
A.~Cichocki, S.~Amari, K.~Siwek, T.~Tanaka, A.~H. Phan, Icalab toolboxes,
  http://www.bsp.brain.riken.jp/ICALAB.

\bibitem{Jiang2012a}
N.~Jiang, J.~L.~G. Vest-Nielsen, S.~Muceli, D.~Farina, {EMG-based simultaneous
  and proportional estimation of wrist/hand kinematics in uni-lateral
  trans-radial amputees.}, Journal of neuroengineering and rehabilitation 9~(1)
  (2012) 42.
\newblock \href {http://dx.doi.org/10.1186/1743-0003-9-42}
  {\path{doi:10.1186/1743-0003-9-42}}.

\bibitem{Clancy2017}
E.~A. Clancy, C.~Martinez-Luna, M.~Wartenberg, C.~Dai, T.~R. Farrell, {Two
  degrees of freedom quasi-static EMG-force at the wrist using a minimum number
  of electrodes}, Journal of Electromyography and Kinesiology 34 (2017) 24--36.
\newblock \href {http://dx.doi.org/10.1016/j.jelekin.2017.03.004}
  {\path{doi:10.1016/j.jelekin.2017.03.004}}.

\bibitem{Muceli2014}
S.~Muceli, N.~Jiang, D.~Farina, {Extracting Signals Robust to Electrode Number
  and Shift for Online Simultaneous and Proportional Myoelectric Control by
  Factorization Algorithms}, IEEE Transactions on Neural Systems and
  Rehabilitation Engineering 22~(3) (2014) 623--633.
\newblock \href {http://dx.doi.org/10.1109/TNSRE.2013.2282898}
  {\path{doi:10.1109/TNSRE.2013.2282898}}.

\bibitem{Muceli2010d}
S.~Muceli, A.~T. Boye, A.~D'Avella, D.~Farina, {Identifying representative
  synergy matrices for describing muscular activation patterns during
  multidirectional reaching in the horizontal plane.}, Journal of
  Neurophysiology 103~(3) (2010) 1532--1542.
\newblock \href {http://dx.doi.org/10.1152/jn.00559.2009}
  {\path{doi:10.1152/jn.00559.2009}}.

\bibitem{Jiang2009}
N.~Jiang, K.~B. Englehart, P.~a. Parker, {Extracting simultaneous and
  proportional neural control information for multiple-dof prostheses from the
  surface electromyographic signal}, IEEE Transactions on Biomedical
  Engineering 56~(4) (2009) 1070--1080.
\newblock \href {http://dx.doi.org/10.1109/TBME.2008.2007967}
  {\path{doi:10.1109/TBME.2008.2007967}}.

\bibitem{Ikeda2000}
S.~Ikeda, K.~Toyama, {Independent component analysis for noisy data - MEG data
  analysis}, Neural Networks 13~(10) (2000) 1063--1074.
\newblock \href {http://dx.doi.org/10.1016/S0893-6080(00)00071-X}
  {\path{doi:10.1016/S0893-6080(00)00071-X}}.

\end{thebibliography}

\appendix
	
	\section{Minimum description length (MDL)}
	\label{appendix:MDL}

	The MDL method for determining the number of synergies is performed by calculating the maximum likelihood estimates of factor loading matrix $\textbf{A}$ and the unique variances diagonal matrix $\bm{\Psi}$ according to the factor analysis model 
	\begin{equation}
	\mathbf{C}=\textbf{AA}^{T}+\bm{\Psi}
	\end{equation}
	 where $\textbf{C}$ is the covariance matrix of $\textbf{M}_{m\times n}$  the multi-channel EMG signal matrix  with $m$ channels and $n$ samples. 
	 
	 This is done for different number of synergies ($r$) between $ 1 \le r \le \frac{1}{2} (2m+1-\sqrt{8m+1})$ in order to minimise the MDL. The boundary for $r$ is set by comparing  the number of equations with unknowns in order to have an algebraic solution for equation \ref{eq_MLE}. 
	
	\begin{equation}\label{eq_MLE}
	L(\mathbf{A},\bm{\Psi})=-\frac{1}{2}\left\lbrace \mathtt{tr}(\mathbf{C}(\bm{\Psi}+\mathbf{AA}^{T})^{-1})+\log(\det(\bm{\Psi}+\mathbf{AA}^{T}))+ m\log 2\pi \right\rbrace 
	\end{equation}
	
	\begin{equation}\label{eq_MDL}
	\mathrm{MDL} =-L(\mathbf{A},\bm{\Psi}) + \frac{\log n}{n} \left( m(r+1)-\frac{r(r-1)}{2}\right) 
	\end{equation}
	The number of synergies $r$ are selected to minimise the MDL value in equation \ref{eq_MDL}.

\end{document}